\documentclass[aps,preprint,superscriptaddress]{revtex4}
\usepackage{amsfonts}
\usepackage{amssymb}
\usepackage{amsmath}
\usepackage{amsfonts}
\usepackage{amssymb}
\usepackage{graphicx}

\begin{document}

\title{Relaxation of A Thermally Bathed Harmonic Oscillator: A Study Based on
the Group-theoretical Formalism}

\author{Yan Gu}
\affiliation{Department of Modern Physics, University of Science and Technology
of China, Hefei 230026, Anhui, China}
\email{ygu@ustc.edu.cn}
\author{Jiao Wang}
\affiliation{Department of Physics and Fujian Provincial Key Laboratory of Low
Dimensional Condensed Matter Physics, Xiamen University, Xiamen 361005, Fujian,
China}
\affiliation{Lanzhou Center for Theoretical Physics, Lanzhou University,
Lanzhou 730000, Gansu, China}

\begin{abstract}
The quantum dynamics of a damped harmonic oscillator has been extensively studied since the 1960s of the last century. Here, with a distinct tool termed the ``group-theoretical characteristic function (GCF)", we investigate analytically how a harmonic oscillator immersed in a thermal environment would relax to its equilibrium state. We assume that the oscillator is at a pure state initially and its evolution is governed by a well-known quantum-optical master equation. Taking advantage of the GCF, the master equation can be transformed into a first-order linear partial differential equation, allowing us to write down its solution explicitly. Based on the solution, it is found that, in clear contrast with the monotonic relaxation process of its classical counterpart, the quantum oscillator may demonstrate some intriguing non-monotonic relaxation characteristics. In particular, when the initial state is a Gaussian state (i.e., a squeezed coherent state), there is a critical value of the environmental temperature below which the entropy will first increase to reach its maximum value, then turn down and converge to its equilibrium value from above. Conversely, when the temperature exceeds the critical value, the entropy converges monotonically to its equilibrium value from below. In contrast, for an initial Fock state, there are two critical temperatures instead and, in between, a new additional phase emerges, where the time curve of entropy features two extreme points. Namely, the entropy will increase to reach its maximum first, then turn down to reach its minimum, from where it begins to increase and converges to the equilibrium value eventually. Other related issues are discussed as well.


\end{abstract}
\maketitle

\section{Introduction}
The evolution of a quantum system that interacts with its environment is of fundamental importance in the theory of open quantum systems \cite{Breuer02, Weiss00}. In this paper, we study the evolution of a quantum harmonic oscillator that moves irreversibly in a thermostatic setting. The reduced density matrix $\mathbf{\rho}$ of the oscillator is assumed to follow the quantum-optical
master equation
\begin{equation}
\label{eq11}
\frac{d\mathbf{\rho}}{dt}=-i\omega[a^{\dag}a,\mathbf{\rho}]+
\Gamma(N_{\beta}+1)(2a\mathbf{\rho}a^{\dag}-\{a^{\dag}a,\mathbf{\rho}\})+
\Gamma N_{\beta}(2a^{\dag}\mathbf{\rho}a-\{aa^{\dag},\mathbf{\rho}\}),
\end{equation}
where $\Gamma$ is the damping constant, $\beta=\frac{1}{k_B T}$ with $k_B$ and $T$ being the Boltzmann constant and the environmental temperature, respectively, $N_{\beta}=(e^{\beta\hbar\omega}-1)^{-1}$ is the expectation number of thermal
quanta, and
\begin{equation}
\label{eq12}
a=\frac{1}{\sqrt{2\hbar M\omega}}(M\omega q+ip),~~
a^{\dag}=\frac{1}{\sqrt{2\hbar M\omega}}(M\omega q-ip).
\end{equation}

Retrospectively, in the early 1970s, Louisell \cite{Louisell73} had derived this master equation to discuss a damped mode of the radiation field in a cavity and was considered by Haake \cite{Haake73} as an example of the generalized master equation constructed by Nakajima and Zwanzig. Since then a large number of works have been devoted to this equation. Today, it plays a central role not only in interpreting various quantum-optical experiments \cite{Walls08, Miguel07}, but also in studying the environment-induced decoherence of open quantum systems \cite{Breuer02}. However, it is also important to note that this equation provides only an approximate description of the motion of a physical oscillator, as has been remarked in the literature (see, e.g., p.17 of Ref. \cite{Carmichael02}). Thus, one should be careful with its applicability in dealing with physical problems.

Although it is generally impossible to solve Eq.~\eqref{eq11} directly to obtain the solution $\rho(t)$ in the operator form, there are various methods to transform $\rho(t)$ into a c-number function (or distribution), in terms of which Eq.~\eqref{eq11} can be transformed into a c-number equation, accordingly. One way is to employ the Glauber-Sudarshan $P$ representation \cite{Glauber63, Sudarshan63} by representing the density operator as an ensemble of coherent states. An alternative approach is to use the Wigner distribution \cite{Wigner32} or a phase-space quasi-probability distribution \cite{Agarwal70} that resembles the classical probability distribution more closely. After replacing the density operator $\rho(t)$ with these quasi-probability distributions, the operator master equation \eqref{eq11} can be converted to a c-number Fokker-Planck equation \cite{Walls08, Miguel07, Carmichael02, Risken89}. A technique for solving the resultant Fokker-Planck equation is to utilize its Fourier transform (i.e., in terms of the characteristic function) \cite{Mises64}, as Wang and Uhlenbeck did in developing the corresponding theory for a classical harmonic oscillator \cite{Wang45}. Notably, the differential equation in terms of the quantum characteristic function is a first-order linear partial differential equation. Its solution can be obtained with standard solving methods \cite{Walls08, Carmichael02, Risken89}.

After solving Eq.~\eqref{eq11}, it is interesting to examine how the system relaxes towards the equilibrium state, a mixed state, starting from a pure quantum state. This problem has been thoroughly studied in the literature as well (see Refs.~\cite{Dodonov00, Valeriano20, Marian93A, Marian93B, Marian13} and references therein). For example, in
Refs. \cite{Dodonov00, Valeriano20}, the authors solved the relevant Fokker-Planck equation for the time-dependent Wigner distribution and demonstrated that, in the low-temperature regime, the evolution of quantum purity exhibits nonmonotonic behavior for initial Gaussian and/or Fock states. In Refs.~\cite{Marian93A, Marian93B, Marian13}, the authors addressed Eq.~\eqref{eq11} using the quantum characteristic function of the density operator and conducted a detailed analysis of relaxation properties -- revealed through purity and other quantum metrics -- for various initial states. It is worth noting that nonmonotonic relaxation behaviors were also identified in these studies.

In a previous publication \cite{Gu85}, one of the authors introduced a group-theoretical framework for quantum mechanics, which postulates that the Fourier transform of a quantum system's density operator must be represented as a non-negative definite function on a group. This c-number function is termed the ``group-theoretical characteristic function (GCF)'' of the density operator, and the group on which it is defined is referred to as a ``phase group'' \cite{Gu85}. Under this Fourier-transform mapping, all observables (self-adjoint operators) of the system are mapped to Hermitian distributions with compact support and convolution multiplication on the group \cite{Gu92, Gu20}, so that the non-commutativity of observables is explicitly encoded in the non-commutativity of group elements, establishing a direct link between algebraic structures and physical phenomena. For systems described by canonical variables, the phase group corresponds to the well-known Heisenberg-Weyl group, reinforcing the formalism's connection to quantum mechanics. Conceptually, this group-theoretical formalism and the associated characteristic function are superior because they align more closely with quantum mechanical principles featuring the non-commutativity of observables. In addition, in applications, this formalism offers a key advantage: it grants us the freedom to specify both the group representation and the coordinates of the group manifold. This, in turn, enables us to tailor the form of the GCF to specific problem requirements. Thus, we can select the most mathematically convenient GCF form for solving a given problem.

In this paper, our motivations are twofold. The first is to review and discuss the group-theoretical formalism in studying the dynamical evolution of open quantum systems, in the hope that it may find more applications in future in view of its distinct advantages. The second is to use the GCF to solve Eq.~\eqref{eq11} and, based on the solutions, identify the features peculiar to the quantum relaxation process of the damped harmonic oscillator in comparison with the classical case. We will show that, indeed, even for such a simple model, the quantum relaxation process may be much more diverse than its classical counterpart. This part is scientifically interesting by itself, which is also an illustration of the effectiveness of the GCF approach in applications.

Mathematically, for the specific problem we are dealing with here, the GCF approach is equivalent to that in Refs. \cite{Marian93A, Marian93B, Marian13}. Here, in this paper, we will provide more and detailed discussions of the quantum relaxation process based on analytical solutions of the master equation \eqref{eq11}.

The organization of the remaining sections is as follows. In Sec.~2, we will first introduce the GCF representation of a system described with canonical variables by making use of two different coordinate systems on the Heisenberg-Weyl group. They are related to the phase-space variables $(\textbf{q,p})$ and the complex variables ($\textbf{a}, \textbf{a}^{\dag}$), respectively. In Sec.~3 and Sec.~4, we present the relaxation phases of the damped harmonic oscillator with an initial Gaussian and an initial Fock state, respectively. The former is a squeezed coherent state and the latter is an energy eigenstate of the isolated harmonic oscillator. As a comparison with their classical counterparts, in Sec.~5 we will discuss the relaxation of the classical harmonic oscillator. Conclusions and discussions are given in Sec.~6. Finally, in Appendix A, the analytical expression for describing the evolution of diagonal elements of the relevant density matrix in the Fock state basis is derived to illustrate the application of the GCF.

\section{GCF representation and quantum-optical master equation}

We start by considering a quantum system with a pair of canonical variables, $(\textbf{q,p})$, and look upon the commutation relation $[\textbf{q,p}]=i\hbar$ as a Lie bracket of a three dimensional Heisenberg-Weyl Lie algebra \cite{Wolf75}. This Lie algebra generates a nilpotent Lie group, denoted as $H(1)$ in the following. Let
\begin{equation}
\label{eq21}
g\rightarrow\textbf{U}(g)=e^{i(x \textbf{p}+y \textbf{q}+z \hbar)},~~g\in H(1)
\end{equation}
be an irreducible unitary representation of $H(1)$ on the state vector space $\mathfrak{H}$ of the system, where
$\{(x,y)\in \mathbb{R}^{2},\hspace{1mm} 0\leq z<2\pi/\hbar $\} represents a canonical coordinate system on the group $H(1)$. Following Ref. \cite{Gu85}, the GCF of the density matrix $\mathbf{\rho}(t)$ of the quantum system at time $t$ is defined as
\begin{equation}
\label{eq22}
\varphi(t,g)\equiv \textrm{Tr}_{\mathfrak{H}}[\mathbf{\rho}(t)\textbf{U}(g)]
= e^{iz \hbar+v(x,y,t)}
\end{equation}
with the inverse mapping
\begin{equation}
\label{eq23}
\mathbf{\rho}(t)=\int_{H(1)}\varphi(t,g)\textbf{U}^{\dag}(g)dg,
\end{equation}
where $dg=(\hbar/2\pi)^2 dxdydz$ is the bi-invariant Haar measure on $H(1)$. If we use the complex variables ($\textbf{a},\textbf{a}^{\dag}$) instead of $(\textbf{q,p})$, the unitary representation of $H(1)$ in $\mathfrak{H}$ has the form $\textbf{U}(g)=e^{u\textbf{a}^{\dag}-\bar{u}\textbf{a}+iz \hbar}$ with $u=u_{1}+iu_{2}=-\sqrt{\frac{\hbar M\omega}{2}}x+i\sqrt{\frac{\hbar} {2 M\omega}}y$, which gives an alternative expression of the GCF
\begin{equation}
\label{eq24}
\varphi(t,g)=\textrm{Tr}_{\mathfrak{H}}[\mathbf{\rho}(t)\textbf{U}(g)]
\equiv e^{iz \hbar+v(u,t)}.
\end{equation}

Using the formulae
\begin{equation}
\label{eq245}
\textbf{aU}(g)=(-\frac{\partial}{\partial \bar{u}}+
\frac{u}{2})\textbf{U}(g), ~~~~\textbf{a}^{\dag}\textbf{U}(g)=
(\frac{\partial}{\partial u}+\frac{\bar{u}}{2})\textbf{U}(g),
\end{equation}
we can convert the quantum-optical master equation \eqref{eq11} into a c-number equation of $v(u,t)$ as
\begin{equation}
\label{eq25}
\frac{\partial v(u,t)}{\partial t} =-\{[(\Gamma u_{1}+\omega u_{2})\frac{\partial}
{\partial u_{1}}+(\Gamma u_{2}-\omega u_{1})\frac{\partial}{\partial u_{2}}]v(u,t)
+(1+2N_{\beta})\Gamma |u|^{2}\},
\end{equation}
or, alternatively, of $v(x,y,t)$ as
\begin{equation}
\label{eq26}
\frac{\partial v(x,y,t)}{\partial t} =[(\frac{y}{M}-\Gamma x)\frac{\partial}
{\partial x}-(M \omega^{2}x+\Gamma y)\frac{\partial}{\partial y}]v(x,y,t)-
\frac{\hbar}{2}(1+2N_{\beta})\Gamma (M \omega x^{2}+\frac{y^{2}}{M\omega}).
\end{equation}

Note that Eqs.~\eqref{eq25} and \eqref{eq26} are first-order linear partial differential equations that can be solved straightforwardly. Meanwhile, as shown in the following, the flexibility for generating different forms of the GCF by adopting different coordinates of the group manifold, and in turn, the different forms of the master equation [see Eqs.~\eqref{eq25} and \eqref{eq26}], can be a remarkable advantage of the group-theoretical formalism in applications.

Moreover, in the GCF representation, the expectation value of an observable, say, $A$, can be conveniently obtained as well via the transformation
\begin{equation}
\label{eq27}
\langle A \rangle=\textrm{Tr}_{\mathfrak{H}}[A\mathbf{\rho}]=\textrm{Tr}_{\mathfrak{H}}
[A \textbf{U}(g)\mathbf{\rho}]|_{g=\textbf{e}},
\end{equation}
where $\textbf{e}$ is the identity element of $H(1)$. Given the solution $v(x,y,t)$ or $v(u,t)$, the expectation value of $A$ at time $t$ can then be written down straightforwardly. For example, for the position $\textbf{q}$,
\begin{equation}
\label{eq28}
\langle \textbf{q} \rangle_t=\textrm{Tr}_{\mathfrak{H}}
[\textbf{q} \textbf{U}(g)\mathbf{\rho}(t)]|_{g=\textbf{e}}=(-i\frac{\partial}{\partial y}-\frac{\hbar}{2}x)e^{v(x,y,t)}|_{x=y=0}.
\end{equation}
When the initial state is Gaussian, $v(0,0,t)=0$ [see Eqs. \eqref{eq33} and \eqref{eq34}], it reduces further to
\begin{equation}
\label{eq29}
\langle \textbf{q} \rangle_t=-i\frac{\partial}{\partial y}{v(x,y,t)}|_{x=y=0}.
\end{equation}
Similarly,
\begin{equation}
\label{eq210}
\langle \textbf{p} \rangle_t=-i\frac{\partial}{\partial x}{v(x,y,t)}|_{x=y=0},
\end{equation}
\begin{equation}
\label{eq211}
\langle \textbf{q}^2 \rangle_t=-\{\frac{\partial^2}{\partial y^2}{v(x,y,t)}+[\frac{\partial}{\partial y}{v(x,y,t)}]^2\}|_{x=y=0},
\end{equation}
\begin{equation}
\label{eq212}
\langle \textbf{p}^2 \rangle_t=-\{\frac{\partial^2}{\partial x^2}{v(x,y,t)}+[\frac{\partial}{\partial x}{v(x,y,t)}]^2\}|_{x=y=0}.
\end{equation}
In terms of $v(u,t)$ and by referring to \eqref{eq245}, we have additionally that
\begin{equation}
\label{eq213}
\langle \textbf{a}^\dag \textbf{a} \rangle_t=\mathfrak{L} e^{v(u,t)}|_{u=0}, ~~~
\langle (\textbf{a}^\dag \textbf{a})^2 \rangle_t=\mathfrak{L}^2 e^{v(u,t)}|_{u=0},
\end{equation}
where the operator $\mathfrak{L}$ reads
\begin{equation}
\label{eq214}
{\mathfrak{L}}=\frac{|u|^{2}}{4}-\frac{1}{4}(\frac{\partial^{2}}{\partial u_{1}^{2}}+
\frac{\partial^{2}}{\partial u_{2}^{2}})-\frac{i}{2}(u_{1}\frac{\partial}
{\partial u_{2}}-u_{2}\frac{\partial}{\partial u_{1}})
-\frac{1}{2}.
\end{equation}
In the following sections, we will use these expressions to discuss in detail how the variances of position and energy evolve as the system relaxes.

\section{Relaxation of an initial Gaussian state}

When the initial state of the system is a Gaussian state, $|\psi_{G}\rangle$, centered at $(\bar{q},\bar{p})$ with variance $\sigma^2$ in $q$ that
\begin{equation}
\label{eq31}
\langle q|\psi_{G}\rangle=(\frac{1}{2\pi\sigma^{2}})^{\frac{1}{4}}
e^{-\frac{(q-\bar{q})^{2}}{4\sigma^{2}}+\frac{i}{\hbar}\bar{p}(q-\bar{q})},
\end{equation}
the GCF in terms of $(x, y, z)$ [see Eq.~\eqref{eq22}] is preferable for solving the corresponding master equation~\eqref{eq26}. For the initial density operator $\rho(0)=|\psi_{G}\rangle \langle\psi_{G}|$, the GCF reads $\varphi(0, g)=\langle\psi_{G}| \textbf{U}(g)|\psi_{G}\rangle=e^{iz \hbar+v(x,y,0)}$, where
\begin{equation}
\label{eq32}
v(x,y,0)=i(x \bar{p}+y \bar{q})-(\frac{\hbar^{2}x^{2}}{8\sigma^{2}}+
\frac{\sigma^{2}y^{2}}{2}).
\end{equation}
It is straightforward to verify that the solution of the master equation \eqref{eq26} with the initial condition \eqref{eq32} can be written down as
\begin{equation}
\label{eq33}
v(x,y,t)=A(t)x+B(t)y+a(t)x^{2}+b(t)y^{2}+c(t)xy,
\end{equation}
where
\begin{equation}
\label{eq34}
\left\{\begin{array}{l}
A(t)=ie^{-\Gamma t}(\bar{p}\cos\omega t-M\bar{q}\omega\sin\omega t),\\
B(t)=ie^{-\Gamma t}(\bar{q}\cos\omega t+\frac{\bar{p}}{M\omega}\sin\omega t),\\
a(t)=-\frac{\hbar M \omega}{4}(2N_{\beta}+1)(1-e^{-2\Gamma t})-\frac{1}{2}
e^{-2\Gamma t}(\frac{\hbar^{2}}{4\sigma^{2}}\cos^{2}\omega t+M^{2}\sigma^{2}
\omega^{2}\sin^{2}\omega t),\\
b(t)=-\frac{\hbar }{4M \omega}(2N_{\beta}+1)(1-e^{-2\Gamma t})-\frac{1}{2}
e^{-2\Gamma t}(\sigma^{2}\cos^{2}\omega t+\frac{\hbar^{2}}{4M^{2}\sigma^{2}
\omega^{2}}\sin^{2}\omega t),\\
c(t)=\frac{e^{-2\Gamma t}}{8M\sigma^{2}\omega}(4M^{2}\sigma^{4}\omega^{2}
-\hbar^{2})\sin2\omega t.
\end{array}\right.
\end{equation}

In the GCF representation, the purity of the system at time $t$ has the form \cite{Gu85}
\begin{equation}
\label{eq35}
P(t)\equiv \textrm{Tr}_{\mathfrak{H}}  [\mathbf{\rho}^2(t)]=\int_{H(1)}|\varphi(t,g)|^2dg=
\frac{\hbar}{2\pi}\int dx\int dy~e^{2 \mathbf{Re}[v(x,y,t)]}
\end{equation}
and in turn, for the initial Gaussian state \eqref{eq31}, the Prigogine entropy of the system is \cite{Prigogine73}
\begin{equation}
\label{eq36}
\begin{array}{l}S(t)\equiv -\ln P(t)=\ln \frac{2}{\hbar}\sqrt{4a(t)b(t)-c(t)^{2}}\\
~~~~~=\frac{1}{2}\ln[e^{-4\Gamma t}+(1-e^{-2\Gamma t})^{2}(2N_{\beta}+1)^{2}\\
~~~~~~~~+2(1-e^{-2\Gamma t})e^{-2\Gamma t}(2N_{\beta}+1)\cosh2r].
\end{array}
\end{equation}
Here, the squeeze factor $r$ is defined as $e^{-r}={\sigma}/{\sigma_{c}}$, with~$\sigma_{c}=\sqrt{\frac{\hbar}{2M\omega}}$. As we will discuss later in Sec.~5, for a classical harmonic oscillator with an initially localized phase-space distribution, the time evolution of its entropy [see Eq.~\eqref{eq59}] depends on neither the bath temperature nor its initial position in the phase space but the damping ratio $\Gamma/(2\omega)$. It is thus perplexing somehow to comprehend such differences between the quantum and the classical dynamical behaviour of the system in the context of non-equilibrium statistical mechanics.

\begin{figure}[t]
\centering
\includegraphics[width=8cm]{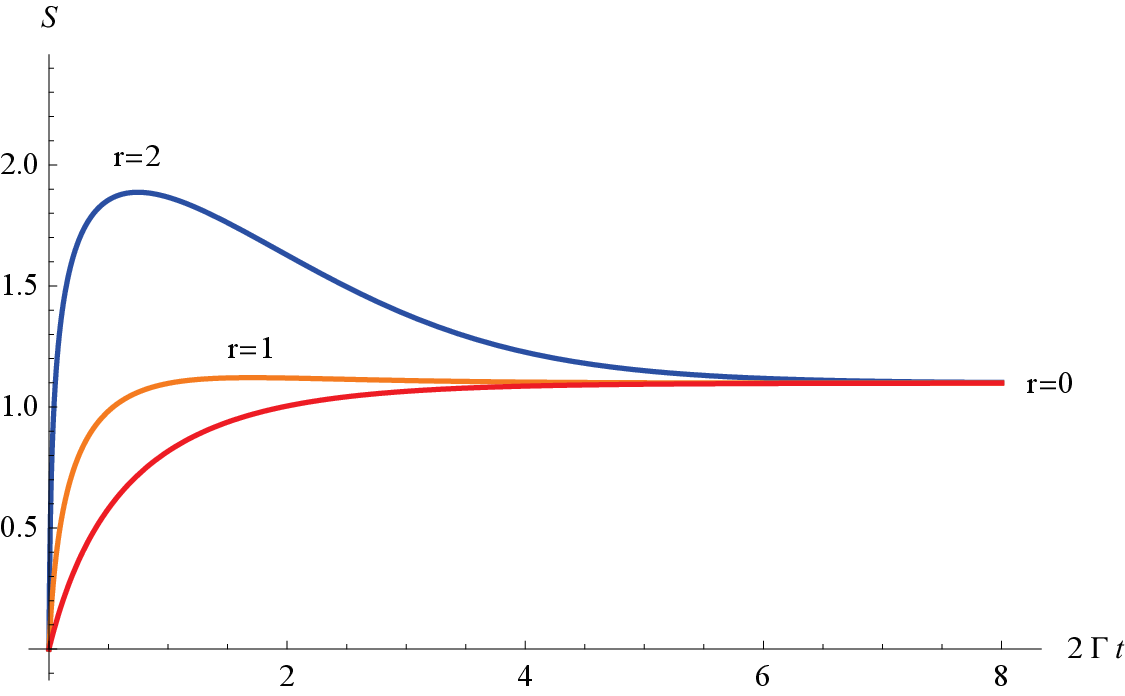}
\caption{The entropy as a function of time, $S(t)$, for an initial Gaussian state. The parameter $N_{\beta}$ is fixed to be $N_{\beta}=1$. From bottom to top, the red curve is for $r=0$ and $N_c=0$, the orange curve is for $r=1$ and $N_{c}\approx 1.38$, and the blue curve is for $r=2$ and $N_{c}\approx 13.15$. Note that $N_{c}=\sinh^{2}r$.}
\label{St-Gaussian}
\end{figure}

To analyze Eq.~\eqref{eq36} further, we first note that initially $S(0)=0$, and in the long time limit $t\to \infty$, $S(t)\to S_{eq}=\ln (2N_{\beta}+1)$, regardless of the squeeze factor $r$ of the initial state. It suggests that the system will relax to the equilibrium state, no matter what initial Gaussian state it starts with. To characterize the ensuing relaxation process, we can use the entropy production rate $R(t)=dS(t)/dt$ as a useful indicator. In particular, $R(0)$ can be used as a measure of the rate at which a pure state deteriorates into a mixture \cite{Zurek93}. From Eq.~\eqref{eq36}, we have
\begin{equation}
\label{eq37}
R(0)=2\Gamma[(2N_{\beta}+1)\cosh2r-1],
\end{equation}
suggesting that the minimum of the initial entropy production rate for a Gaussian state corresponds to the coherent state with $r=0$. In other words, based on the quantum-optical master equation, the most stable Gaussian state for a thermally bathed harmonic oscillator is the coherent state. For this special case ($r=0$), the entropy reads
\begin{equation}
\label{eq38}
S(t)=\ln[1+2N_{\beta}(1-e^{-2\Gamma t})],
\end{equation}
from which we can tell that the entropy of a coherent state will increase monotonically and saturate to its equilibrium value $S_{eq}$ (see the red curve in Fig.~\ref{St-Gaussian}).

For an initial state with $r\neq 0$, the behaviour of the entropy $S(t)$ becomes more complicated. To characterize $S(t)$, we consider the roots of $R(t)=0$. It follows that for $N_{\beta}<N_{c}\equiv \sinh^{2}r$, the only finite solution of $R(t)=0$ is
\begin{equation}
\label{eq39}
t_{m}=\frac{1}{2\Gamma}\ln[\frac{2+4N_{\beta}+4N_{\beta}^{2}-2(1+2N_{\beta})
\cosh2r}{(1+2N_{\beta})(1+2N_{\beta}-\cosh2r)}],
\end{equation}
when $S(t)$ adopts its maximum value
\begin{equation}
\label{eq310}
S_{max}=S(t_m)=\frac{1}{2}\ln\frac{(1+2N_{\beta})^{2}(\cosh^{2}2r-1)}
{2[(1+2N_{\beta})(\cosh2r-1)-2N_{\beta}^{2}]}.
\end{equation}
Note that $S_{max}$ is greater than $S_{eq}$ and increases monotonically as $|r|$. As a result, the time curve of the entropy features a ``hump'': The entropy will increase monotonically before $t_m$, then turn down to converge to $S_{eq}$ from above monotonically (see, e.g., the blue curve in Fig.~\ref{St-Gaussian}).

Because $t_{m}$ goes from $\frac{\ln2}{2\Gamma}$ to $\infty$ as $N_{\beta}$ increases from 0 to $N_{c}$, such a hump feature of $S(t)$ only shows up when $N_{\beta}<N_{c}$. For $N_{\beta}>N_{c}$, $S(t)$ will converge to $S_{eq}$ monotonically from below. Therefore, $N_c$ serves as a critical value that separates these two ``phases'' of the relaxation behavior that own qualitatively distinct characteristics. Note that $N_{\beta}=1/(e^{\hbar\beta\omega}-1)$ and $\beta=1/(k_B T)$, we can determine from $N_{\beta}=N_c$ the critical environmental temperature $T_c$ that separates
these two phases physically: $T_c=\hbar\omega/[k_B\ln(1+1/\sinh^{2}r)]$. For $N_{\beta}<N_c$, we have $T<T_c$.

Now let us turn to the $q$-variance $(\Delta q)^{2}=\textrm{Tr}_{\mathfrak{H}} [\mathbf{\rho}
\mathbf{q}^{2}]-\textrm{Tr}_{\mathfrak{H}} [\mathbf{\rho}\mathbf{q}]^{2}$ to investigate how the quantum state spreads in the $q$ space. Taking the GCF representation \eqref{eq29} and \eqref{eq211} and substituting Eq.~\eqref{eq33}, we have
\begin{equation}
\label{eq311}
(\Delta q)^{2}_{t}=-\frac{\partial^{2} v(x,y,t)}{\partial y^{2}}|_{x=y=0}=-2b(t).
\end{equation}
[See Eq.~\eqref{eq34} for $b(t)$.] By introducing dimensionless variables
$V=(\Delta q)^{2}/\sigma_{c}^{2}$ and $\gamma=\Gamma/\omega$, we can effectively separate $(\Delta q)^{2}_{t}$ into two parts, i.e.,
\begin{equation}
\label{eq312}
V=V_{1}(N_{\beta})+V_{2}(r,\gamma)
\end{equation}
with
\begin{equation}
\label{eq313}
V_{1}=(1-e^{-2\Gamma t})(1+2N_{\beta})
\end{equation}
and
\begin{equation}
\label{eq314}
V_{2}=e^{-2\Gamma t}[e^{-2r}\cos^{2}\frac{\Gamma t}{\gamma}
+e^{2r}\sin^{2}\frac{\Gamma t}{\gamma}].
\end{equation}
Thus, we have $\langle (\Delta q)^{2}\rangle_{t=0}=\sigma^{2}$ and $\langle (\Delta q)^{2}\rangle_{t\rightarrow\infty}=(1+2N_{\beta})\sigma_{c}^{2}$. Note that the squeeze factor $r$ of the initial Gaussian state is contained in $V_2$ only, so $V_1$ reflects a common spreading behavior followed by all Gaussian states and $V_2$ reflects the deviation from it for a given initial Gaussian state. Indeed, in the long-time limit, $V_2\to 0$ and the $q$-variance will be governed by $V_1$ exclusively.

For the particular case of $r=0$ corresponding to an initial coherent state, the $q$-variance evolves as
\begin{equation}
\label{eq315}
\langle (\Delta q)^{2}\rangle_{t}=[1+2N_{\beta}(1-e^{-2\Gamma t})]\sigma_{c}^{2},
\end{equation}
which is non-decreasing, just as that for the classical harmonic oscillator thermally bathed. As the r.h.s of Eq.~\eqref{eq315} does not depend on $\gamma$, the evolution of the $q$-variance is even simpler than its classical counterpart (see Sec.~5).

\begin{figure}[t]
\centering
\includegraphics[width=8cm]{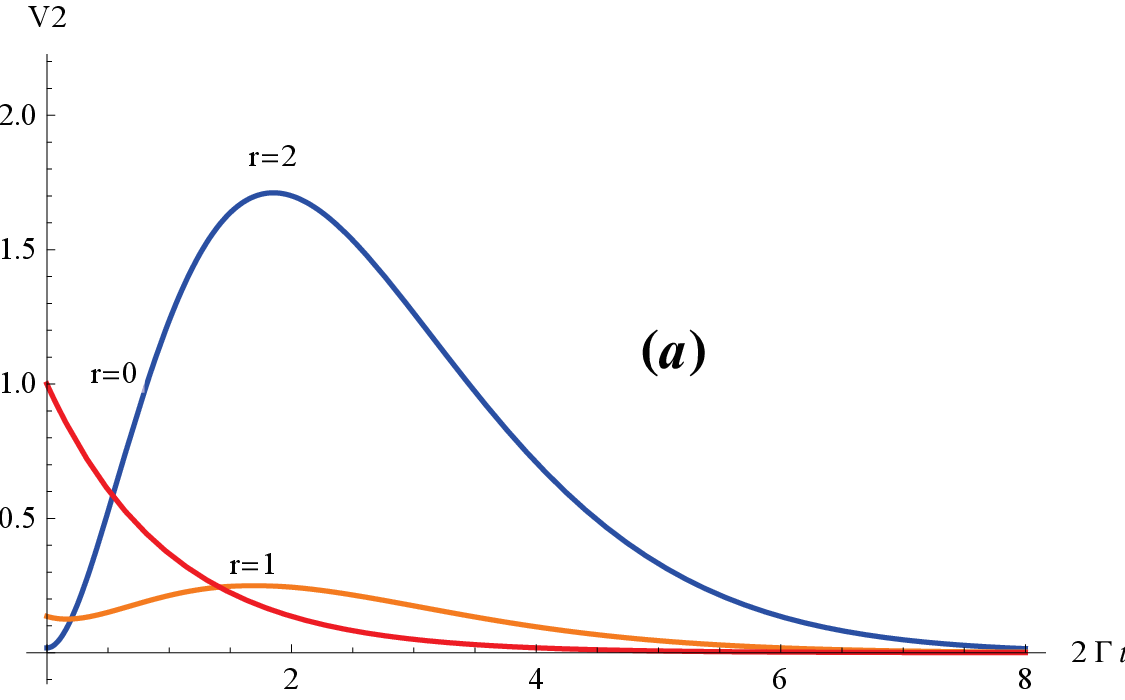}
\includegraphics[width=8cm]{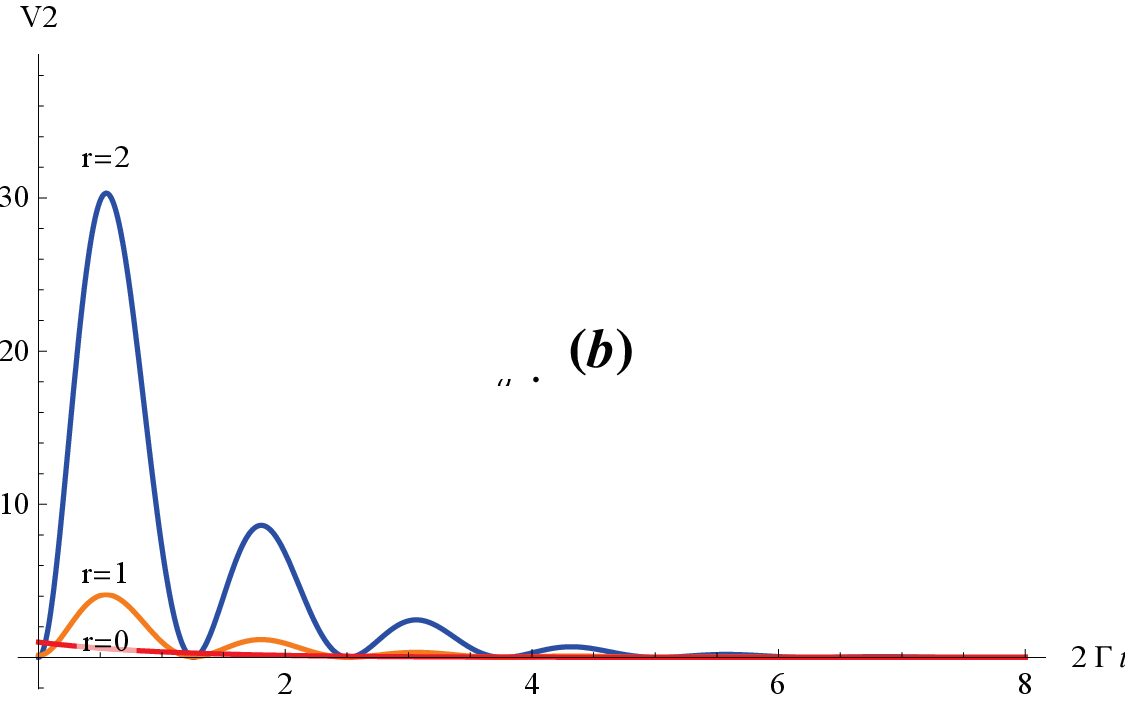}
\caption{Time function $V_{2}$ for an initial Gaussian state with $\gamma =2$ (a) and $\gamma = 0.2$ (b). In both panels, the red line is for $r=0$ and $\gamma_c=0$, the orange line is for $r=1$ and $\gamma_c\approx 3.626$, and the blue line if for $r=2$ and $\gamma_c\approx 27.29$.}
\label{V2-Gaussian}
\end{figure}

For $r\neq 0$, the contribution of $V_{1}$ to the $q$-variance is the same as in the case of $r=0$, but the oscillating term contained in $V_{2}$ will show up given that $\gamma <\gamma_{c}\equiv \sinh2|r|$. Similarly, zeros of the function
\begin{equation}
\label{eq316}
DV_{2}(t)=\frac{dV_{2}}{dt}=-\frac{2\Gamma}{\gamma}e^{-2(r+\Gamma t)}
[\gamma \cos^{2}\frac{\Gamma t}{\gamma}-(e^{4r}-1)\cos\frac{\Gamma t}{\gamma}
\sin\frac{\Gamma t}{\gamma}+e^{4r}\gamma \sin^{2}\frac{\Gamma t}{\gamma}]
\end{equation}
provide important information of $V_{2}$. Suppose that $DV_{2}(t_m)=0$ and let
$\vartheta_{m}=\Gamma t_{m}/\gamma$, then equation $DV_{2}(t_{m})=0$ can be
rewritten as
\begin{equation}
\label{eq317}
\gamma (\cot\vartheta_{m}+e^{4r}\tan\vartheta_{m})=e^{4r}-1,
\end{equation}
which has real solutions $\vartheta_{m}$ if and only if $\gamma <\gamma_c
=\sinh2|r|$. Otherwise, $V_2$ is a monotonically decreasing function of time. Thus we may regard $\gamma_c$ as a critical damping value depending on the squeeze factor $r$ of the initial Gaussian state.

In Fig.~\ref{V2-Gaussian} the time curves of $V_{2}(r,\gamma)$ for various values of $r$ and $\gamma$ are shown, where the characteristic behavior of $V_2$ and thus of the $q$-variance can be well appreciated. Note that for $\gamma <\gamma_c$, $V_2$ may have an increasing number of extremes as $\gamma/\gamma_c$ decreases, implying that the $q$-variance can be highly nonmonotonic in time.

\section{Relaxation of an initial Fock state}

When the initial state of the system is a Fock state, that is, an energy eigenstate, $|n\rangle$, of the isolated harmonic oscillator system, it is preferable to take the complex coordinate system on $H(1)$ instead for solving the master equation.  The corresponding GCF is
\begin{equation}
\label{eq41}
\varphi(0,g)|=\langle n|\textbf{U}(g)|n \rangle =e^{iz \hbar
+\frac{|u|^{2}}{2}}\langle n|e^{-\bar{u}\textbf{a}}e^{u\textbf{a}^{\dag}}
|n \rangle=e^{iz\hbar+v_{n}(u)}.
\end{equation}
Note that, for the coherent state $|\alpha\rangle$, we have
$\langle n|\alpha\rangle=\frac{\alpha^{n}}{\sqrt{n!}}
e^{-\frac{|\alpha|^{2}}{2}}$,
hence
$$\langle n|e^{-\bar{u}\textbf{a}}e^{u\textbf{a}^{\dag}}|n \rangle=
\frac{1}{\pi n!}\int d^{2}\alpha |\alpha|^{2n}e^{-|\alpha|^{2}
+u\bar{\alpha}-\bar{u}\alpha}~~~~~~~~~~~~~~\hspace{5cm}$$
$$~~~~~~=\frac{e^{-|u|^{2}}}{\pi n!}\sum^{n}_{m=0}\binom{n}{m}
\int d\alpha_{1}\alpha_{1}^{2m}e^{-(\alpha_{1}-iu_{2})^{2}}
\int d\alpha_{2}\alpha_{2}^{2(n-m)}e^{-(\alpha_{2}+iu_{1})^{2}}.$$
Using the formula
$\int^{\infty}_{-\infty}dx x^{n}e^{-(x-y)^{2}}
=\frac{\sqrt{\pi}}{(2i)^{n}}H_{n}(iy)$,
we obtain that
\begin{equation}
\label{eq42}
v_{n}(u)\equiv v(u,0)=\ln[\frac{(-1)^{n}}{n!~2^{2n}}e^{-\frac{1}{2}|u|^{2}}
\sum^{n}_{m=0}\binom{n}{m}H_{2m}(u_{2})H_{2(n-m)}(u_{1})],
\end{equation}
where $H_{n}(x)$ is the Hermite polynomial of degree $n$.

Now, we can write down the solution of the master equation \eqref{eq25} with the initial condition \eqref{eq42} as
\begin{equation}
\label{eq43}
v(u,t)=v_{n}(e^{-\Gamma t}u)+\frac{1}{2}(1+2N_{\beta})(e^{-2\Gamma t}-1)|u|^{2}
\end{equation}
and the entropy of the system as
$$S(t)=-\ln[\frac{1}{\pi}\int du_{1}\int du_{2}~e^{2 \mathbf{Re}[v(u,t)]}].$$
Making use of Eq.~\eqref{eq43} and letting $a^{2}=1+(1+2N_{\beta})(e^{2\Gamma t}-1)$, the entropy of the system (with $\rho(0)=|n\rangle\langle n|$) can be expressed further as
\begin{equation}
\label{eq44}
S_{n}(t)=-\ln[\frac{e^{2\Gamma t}}{(n!~2^{2n})^{2}\pi}\sum^{n}_{m_{1},m_{2}}
\binom{n}{m_{1}}\binom{n}{m_{2}}
IP(m_{1},m_{2})IP(n-m_{1},n-m_{2})],
\end{equation}
where \cite{Gradshteyn80}
\begin{eqnarray}
\label{eq45}
IP(m_{1},m_{2})=\int dx e^{-a^{2}x}H_{2m_{1}}(x)H_{2m_{2}}(x)
~~~~~~~~~~~~~~~~~~~~~~~~~~~~~~\nonumber\\
=2^{2(m_{1}+m_{2})}\Gamma(m_{1}+m_{2}+\frac{1}{2})\frac{(1-a^{2})^{m_{1}+m_{2}}}
{a^{2(m_{1}+m_{2})+1}}~~~~~~~~~~~~~\nonumber\\
~_{2}F_{1}(-2m_{1},-2m_{2},\frac{1}{2}-m_{1}-m_{2},
\frac{a^{2}}{2(a^{2}-1)}).~~~~~~~~
\end{eqnarray}

From Eqs.~\eqref{eq44} and \eqref{eq45}, we have
\begin{equation}
\label{eq46}
S_{n}(0)=0,~S_{n}(t)_{t\rightarrow \infty}=S_{eq}=\ln(1+2N_{\beta}),
\end{equation}
the same as in the case of an initial Gaussian state as expected, and
\begin{equation}
\label{eq47}
\frac{dS_{n}}{dt}|_{t=0}=4\Gamma[(2n+1)N_{\beta}+n].
\end{equation}
We also note that the expression of $S_{0}(t)$ given by Eq.~\eqref{eq44} is the same as that given by Eq.~\eqref{eq38} for the coherent state, because the ground state $|0\rangle$ of an isolated harmonic oscillator is exactly the coherent state itself.

In general, the expression of $S_{n}(t)$ becomes increasingly complicated as $n$ increases. However, we find that $S_{n}(t)$ exhibits identical qualitative characteristics for all $n$. For this reason, we will focus on $S_{1}(t)$ as a representative example in the following. It reads
\begin{equation}
\label{eq48}
S_{1}(t)=\ln[ \frac{[e^{-2\Gamma t}+(1-e^{-2\Gamma t})(1+2N_{\beta})]^{3}}
{e^{-4\Gamma t}+(1-e^{-2\Gamma t})^{2}(1+2N_{\beta})^{2}}].
\end{equation}

To decode how bath temperature $T$ affects $S_{1}(t)$, we check again the extreme points of $S_{1}(t)$ and see how they depend on $N_{\beta}$. Let $R_{1}(t)=dS_{1}(t)/dt$, we find the two roots of $R_{1}(t)=0$ are
\begin{equation}
\label{eq49}
t_{\pm}=\frac{1}{2\Gamma}\ln[\frac{1+2N_{\beta}-2N_{\beta}^{2}-4N_{\beta}^{3}
\pm\sqrt{1+6N_{\beta}+10 N_{\beta}^{2}-8N_{\beta}^{4}}}{(N_{\beta}-1)
(1+2N_{\beta})^{2}}].
\end{equation}
So the necessary condition for $S_{1}(t)$ having extreme points is $N_{\beta} <N_c(1)\equiv 1+d_{1}$, where $d_{1}=\frac{1}{2}(\sqrt{3}-1)\approx 0.366$. So $S_{1}(t)$ is non-decreasing only when $N_{\beta}>N_c(1)$. For $1<N_{\beta} <N_{c}(1)$, there are two extreme points located at $t_{-}$ and $t_{+}$, respectively, satisfying $S(t_{-})>S_{eq}>S(t_{+})$. As $N_{\beta}$ decreases from $N_{c}(1)$ to 1, $t_{+}$ goes to infinity, hence $S_{1}(t)$ features a single maximum at $t_{-}$ when $N_{c}(1)<1$.

\begin{figure}[t!]
\centering
\includegraphics[width=8cm]{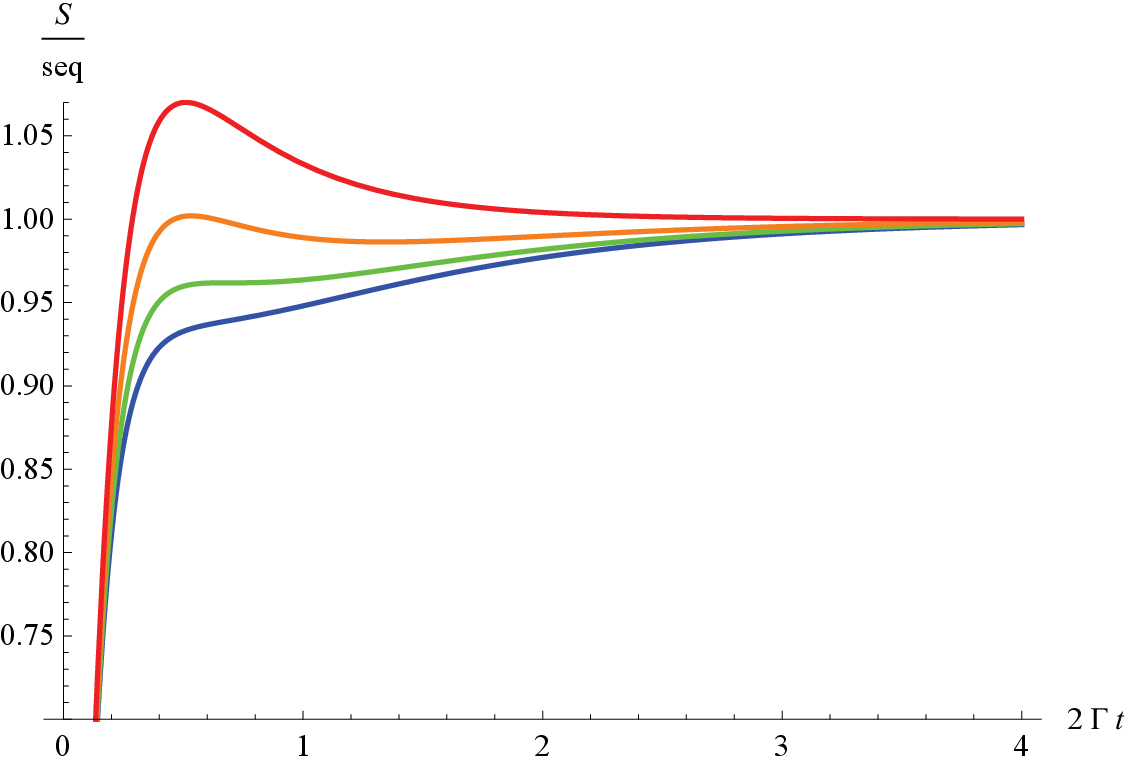}
\caption{Entropy $S_{1}(t)$ for the initial Fock state $|1\rangle$.
From top to bottom, $N_{\beta}=1.0$ (red), $N_{\beta}=1.2$ (orange),
$N_{\beta}=N_c(1)\approx1.366$ (green), and $N_{\beta}=1.5$ (blue).
Note that the red and the green are for the two critical cases, in
between $S(t)$ curve features two extreme points (e.g., the orange curve).}
\label{St-Fock}
\end{figure}
\begin{figure}[t!]
\centering
\includegraphics[width=8cm]{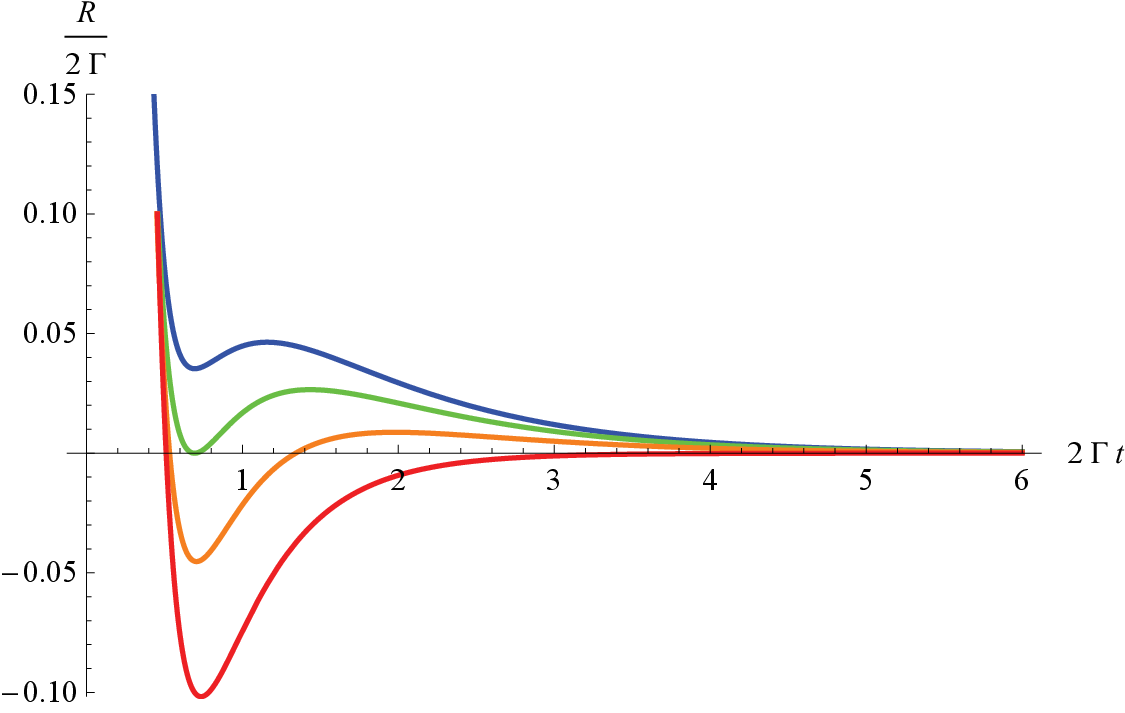}
\caption{The same as Fig.~\ref{St-Fock} but for $R_1 (t)$ instead.}
\label{Rt-Fock}
\end{figure}

In short, compared to the case where the initial state is Gaussian, there is an additional phase of $S(t)$ for an initial Fock state. That is, when the environmental temperature satisfies $T_1<T<T_c(1)$, the curve $S(t)$ features two extreme points, where $T_1$ and $T_c(1)$ correspond to $N_\beta=1$ and $N_\beta=N_c(1)$, respectively. For $T<T_1$, $S(t)$ follows a hump, while for $T>T_c(1)$, $S(t)$ converges to $S_{eq}$ monotonically from below. In Figs.~\ref{St-Fock} and \ref{Rt-Fock}, $S(t)$ and $R(t)$ are plotted, respectively, for four values of $N_\beta$. The three-phase characteristics can be recognized straightforwardly.

Interestingly, it seems that this three-phase characteristic is shared by any Fock state $|n\rangle$, regardless of how large $n$ is. Now, the two critical temperatures correspond to $N_\beta=n$ and $N_\beta=N_c(n)$ instead, where $N_{c}(n)=n+d_{n}$ with $d_{n}$ being a positive number increasing with $n$. Specifically, numerical computation suggests that $d_{2}\approx 0.645$ and $d_{3}\approx 0.924$, respectively. Again, for $N_{\beta}>N_{c}(n)$ and $N_{\beta}<n$, we have the monotonic converging phase and the hump phase, respectively, while for $n<N_{\beta}<N_{c}(n)$, $S_{n}(t)$ is dominated by the two extreme points. When $N_{\beta}$ decreases from $N_{c}(n)$ to $n$, the time $t_{+}$, corresponding to the right minimum point, tends to infinity.

\begin{figure}[t]
\centering
\includegraphics[width=8cm]{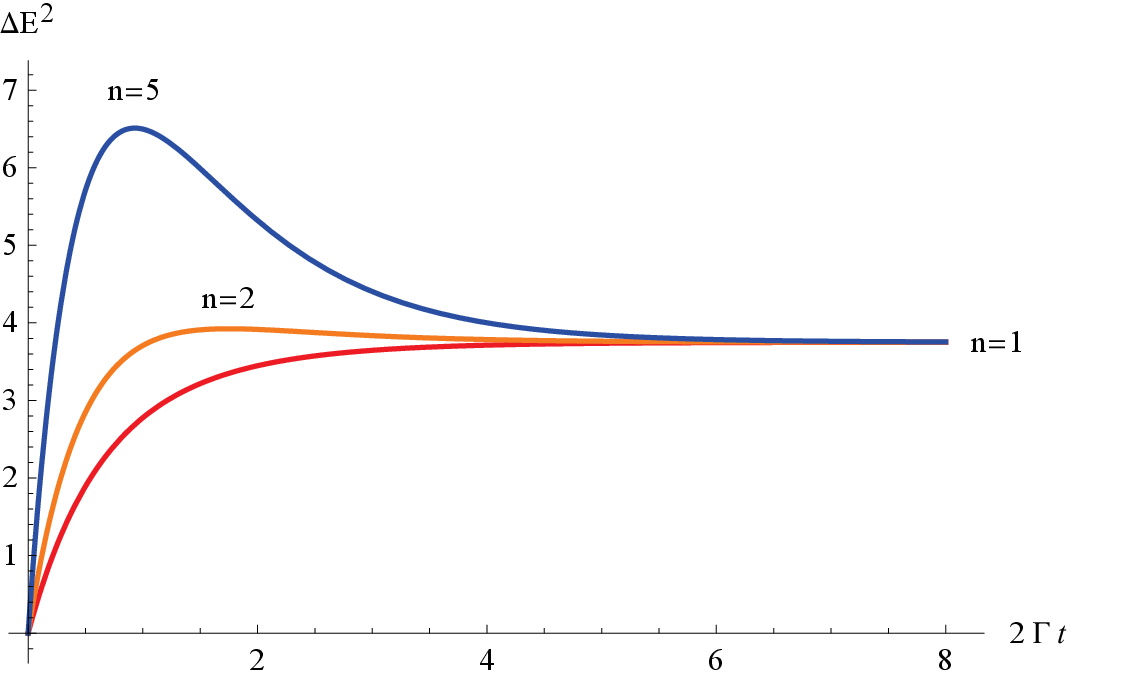}
\caption{Energy variance  $\langle(\Delta E)^{2}\rangle$ as a function
of time for $N_{\beta}=1.5$. The three curves, from bottom to top, are
for $n=1$ (red), $n=2$ (orange), and $n=5$ (blue), respectively.}
\label{VE-Fock}
\end{figure}

It is also interesting to see how the energy spreads in the energy space. To this end, it is convenient to consider the energy variance. Note that in the GCF representation, by substituting the solution $v(u,t)$ given by Eq.~\eqref{eq43} into Eq. \eqref{eq213}, we have
\begin{equation}
\label{eq410}
\langle \textbf{a}^\dag \textbf{a} \rangle_t=N_{\beta}+(n-N_{\beta})e^{-2\Gamma t}
\end{equation}
and
\begin{align}
\label{eq411}
\langle (\textbf{a}^\dag \textbf{a})^2 \rangle_t&=N_\beta(2N_\beta+1)+[4N_\beta(n-N_\beta)+N_\beta+n]e^{-2\Gamma t}\nonumber\\
&+[2N_\beta(N_\beta-2n)+n(n-1)]e^{-4\Gamma t}.
\end{align}
Therefore, by introducing the dimensionless energy variable $E=\textbf{a}^{\dag}\textbf{a}+\frac{1}{2}$, its variance follows
\begin{equation}
\label{eq412}
\langle(\Delta E)^{2}\rangle=(1-e^{-2\Gamma t})[N_{\beta}(1+N_{\beta})
-e^{-2\Gamma t}(N_{\beta}^{2}-2n N_{\beta}-n)].
\end{equation}
Furthermore, since the root of ${d\langle(\Delta E)^{2}\rangle}/{dt}=0$ is
\begin{equation}
\label{eq413}
t_{m}=\frac{1}{2\Gamma}\ln[1+\frac{n(1+2N_{\beta})+N_{\beta}}{(n-N_{\beta})
(1+2N_{\beta})}],
\end{equation}
we conclude that there are two phases of the time curve of $\langle(\Delta E)^{2}\rangle$:
for $N_{\beta}>n$, it has no extreme points and therefore increases monotonically from 0 before saturating to its equilibrium value $\langle(\Delta E)^{2} \rangle_{eq} = N_{\beta}(1+N_{\beta})$; whereas for $N_{\beta}<n$, it features a hump: it increases from 0 to its maximum
\begin{equation}
\label{eq414}
\langle(\Delta E)^{2}\rangle_{max}=
\frac{(n(1+2N_{\beta})+N_{\beta})^{2}}
{4(n(1+2N_{\beta})-N_{\beta}^{2})}
\end{equation}
first, then decays monotonically to $\langle(\Delta E)^{2}\rangle_{eq}$. The time curves of $\langle(\Delta E)^{2}\rangle$ for three different values of $n$ are plotted in Fig.~\ref{VE-Fock} as an illustration.

\section{Relaxation of the classical damped harmonic oscillator}

In order to better appreciate the peculiarity due to the quantum effects in the relaxation process of the quantum harmonic oscillator, here we investigate the relaxation process of its classical counterpart as a comparative study. We assume that for a classical damped harmonic oscillator, its relaxation towards the thermal equilibrium is a Gaussian random process and the evolution of its probability distribution $\rho(q,p,t)$ in phase space $(q,p)$ at time $t$ is governed by the Fokker-Planck equation (see Eq.~(51) in Ref. \cite{Wang45})
\begin{equation}
\label{eq51}
\frac{d\rho}{dt}=\{-\frac{p}{M}\frac{\partial}{\partial q}+\frac{\partial}
{\partial p}[M \omega^{2}q+\Gamma(p+k_B T M\frac{\partial}{\partial p})]\}
\rho.
\end{equation}
Parallel to our discussion in the quantum case, by taking advantage of the classical characteristic function of $\rho(q,p,t)$ defined as
\begin{equation}
\label{eq52}
\varphi(x,y,t)=\int dq\int dp~ e^{i(xp+qy)}\rho(q,p)~\equiv ~e^{v(x,y,t)},
\end{equation}
the Fokker-Planck equation \eqref{eq51} can be transformed into a first-order inhomogeneous linear partial differential equation of $v(x,y,t)$,
\begin{equation}
\label{eq53}
\frac{\partial v(x,y,t)}{\partial t} =[(\frac{y}{M}-\Gamma x)\frac{\partial}
{\partial x}-M \omega^{2}x\frac{\partial}{\partial y}]v(x,y,t)-\Gamma k_B T
M x^{2}.
\end{equation}
For the initially localized probability distribution $\rho(q,p,0)=\delta(q-\bar{q})\delta(p-\bar{p})$,
\begin{equation}
\label{eq54}
v(x,y,0)=i(\bar{p}x+\bar{q}y)
\end{equation}
and the solution of Eq.~\eqref{eq53} can be written down as
\begin{equation}
\label{eq55}
v(x,y;t)=\tilde A(t)x+\tilde B(t)y+\tilde a(t)x^{2}+\tilde b(t)y^{2}+\tilde
c(t)xy,
\end{equation}
where
\begin{equation}
\label{eq56}
\left\{\begin{array}{l}
\tilde A(t)=\frac{ie^{-\frac{\Gamma t}{2}}}{2\alpha^{2}}[2\bar{p}(\alpha^{2}
\cosh\frac{\alpha\Gamma t}{2}-\alpha\sinh\frac{\alpha\Gamma t}{2})+M\Gamma\bar{q}
(\alpha^{2}-1)\alpha\sinh\frac{\alpha\Gamma t}{2}],\\
\tilde B(t)=\frac{ie^{-\frac{\Gamma t}{2}}}{M\Gamma\alpha^{2}}[2\bar{p}\alpha
\sinh\frac{\alpha\Gamma t}{2}+M\Gamma\bar{q}(\alpha^{2}\cosh\frac{\alpha\Gamma
t}{2}+\alpha\sinh\frac{\alpha\Gamma t}{2})],\\
\tilde a(t)=-\frac{Mk_B T e^{-\Gamma t}}{2\alpha^{2}}[\alpha^{2}(e^{\Gamma t}-1)
+1-\cosh\alpha\Gamma t+\alpha\sinh\alpha\Gamma t],\\
\tilde b(t)=-\frac{2 e^{-\Gamma t}k_B T}{ M\Gamma^{2}\alpha^{2}(1-\alpha^{2})}
[\alpha^{2}(e^{\Gamma t}-1)+1-\cosh\alpha\Gamma t-\alpha\sinh\alpha\Gamma t],\\
\tilde c(t)=-\frac{2e^{-\Gamma t}k_B T}{\Gamma\alpha^{2}}(\cosh\alpha\Gamma t-1).
\end{array}\right.
\end{equation}
Note that in Eq.~\eqref{eq56}, we have used the dimensionless parameter $\alpha
\equiv \sqrt{1-\frac{4\omega^{2}}{\Gamma^{2}}}$ instead of the frequency $\omega$
for convenience. Thus, the oscillator is underdamped when $\alpha$ is imaginary
and overdamped when $\alpha$ is real ($0<\alpha\leq 1$).

\begin{figure}[t]
\centering
\includegraphics[width=8cm]{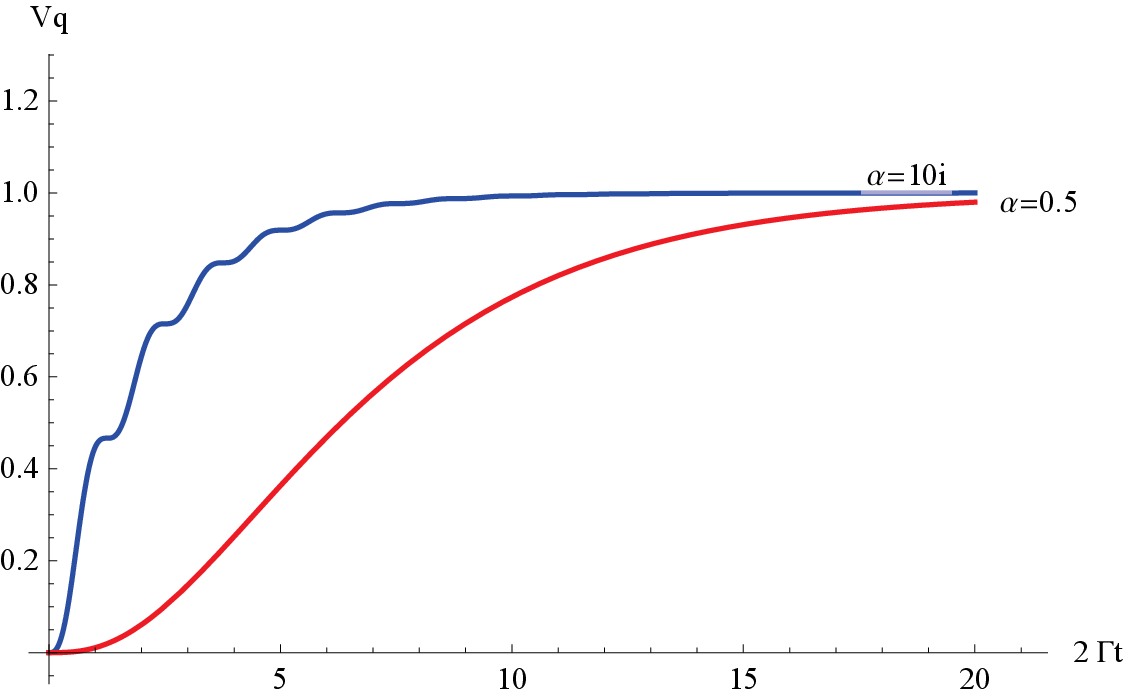}
\caption{Time dependence of the position variance of the classical damped
harmonic oscillator with a localized initial condition. The bottom red curve
is for an overdamped case with a real $\alpha$ ($0.5$) and the top blue curve
is for an underdamped case with an imaginal $\alpha$ $(10i)$, respectively.}
\label{Vq-classical}
\end{figure}

In terms of the characteristic function, the position variance of the oscillator
reads
\begin{equation}
\label{eq57}
\langle (\Delta q)^{2}\rangle_{t}=-2\tilde b(t)=\frac{4k_B T}{M\Gamma^2
\alpha^2(1-\alpha^2)}[\alpha^2+e^{-\Gamma t}(1-\alpha^2-\cosh \alpha
\Gamma t-\alpha\sinh\alpha\Gamma t)],
\end{equation}
which is seemingly the same as Eq.~\eqref{eq311} for the quantum case by simply replacing $b(t)$ with $\tilde b(t)$. It follows that
$\langle (\Delta q)^{2}\rangle_{0}=0$ and $\langle (\Delta q)^{2}\rangle_{eq} =\frac{k_B T}{M\omega^2}$. Furthermore, as
$$\frac{d\langle (\Delta q)^{2}\rangle_{t}}{dt}=\frac{4k_B Te^{-\Gamma t}(\cosh
\alpha\Gamma t-1)}{\alpha^2M\Gamma} \geq 0,$$
the classical $q$-variance is non-decreasing for the localized initial condition discussed here. In Fig.~\ref{Vq-classical}, $Vq\equiv \langle (\Delta q)^{2} \rangle_{t}/\langle (\Delta q)^{2}\rangle_{eq}$ is adopted for showing two examples for the overdamped case ($\alpha=0.5$) and the underdamped case ($\alpha=10i$), respectively.

In contrast with the monotonic behavior of $q$-variance, the evolution of energy variance can be much more complicated. In fact, the $E$-variance $\langle (\Delta E)^{2}\rangle=\langle E^{2}\rangle-\langle E\rangle^{2}$ can be worked out by making use of the following formulae:
$$\langle E\rangle_{t}=\mathfrak{L}_c [e^{v(x,y,t)}]|_{x=y=0},
~~\langle E^{2}\rangle_{t}=\mathfrak{L}_c ^{2}[e^{v(x,y;t)}]|_{x=y=0}, $$
where $\mathfrak{L}_c =-[\frac{1}{2M}\frac{\partial^{2}}{\partial x^{2}} +\frac{M \omega^{2}}{2}\frac{\partial^{2}}{\partial y^{2}}]$ is a differential operator on the characteristic function.

\begin{figure}[t]
\centering
\includegraphics[width=8cm]{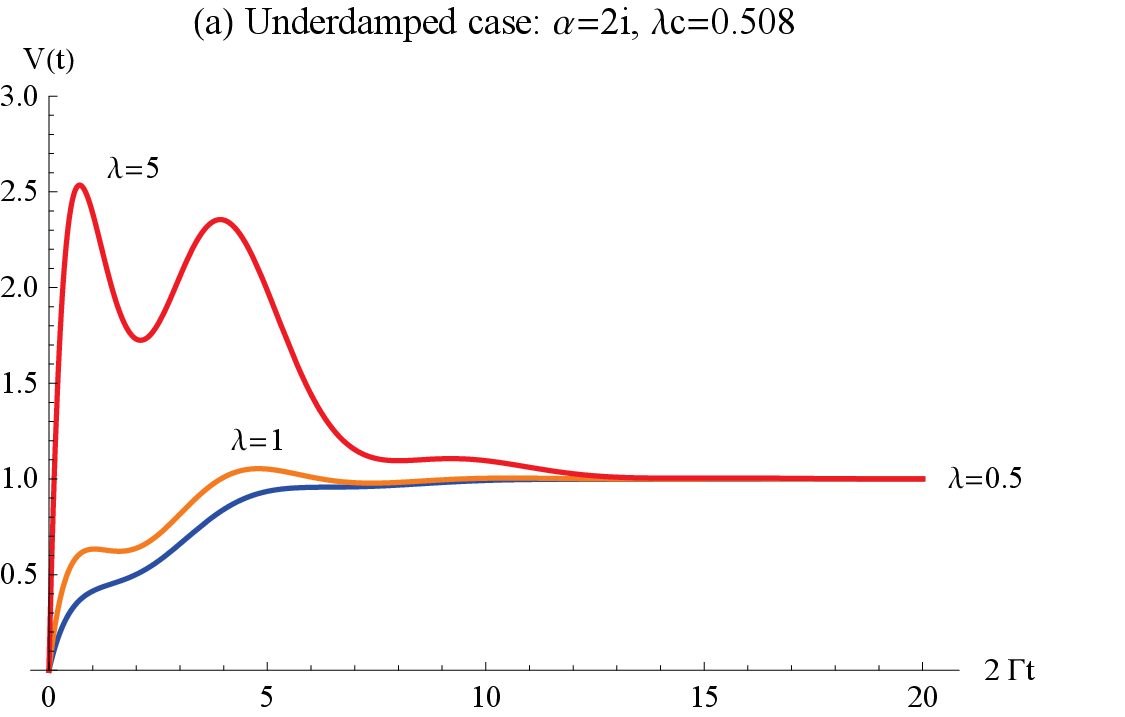}
\includegraphics[width=8cm]{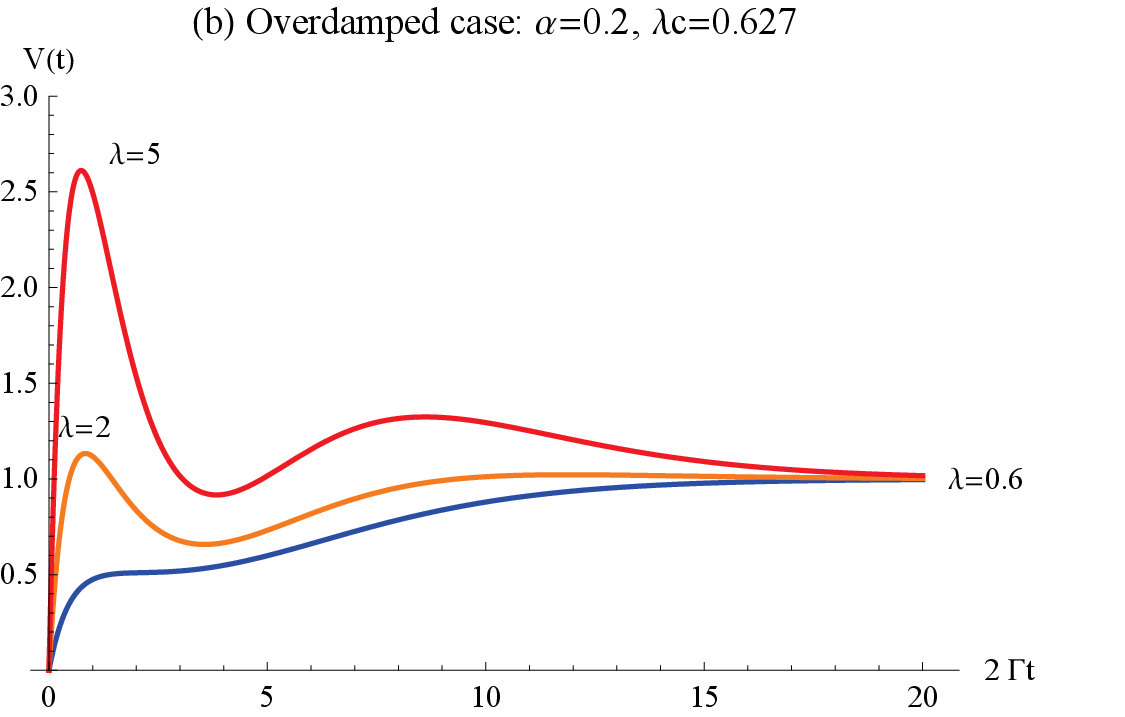}
\caption{Time dependence of the energy variance of the classical damped
harmonic oscillator for the underdamped case with $\alpha=2i$ and $\lambda_{c}
\approx 0.508$ (a) and the overdamped case with $\alpha=0.2$ and $\lambda_{c}
\approx 0.627$ (b). In (a), from bottom to top, the three curves are for initial
energy $E_{0}=\lambda k_B T$ with $\lambda=0.5$ (blue), $\lambda=1$ (orange),
and $\lambda=5$ (red), respectively, while in (b), from bottom to top, the three
curves are for $\lambda=0.6$ (blue), $\lambda=2$ (orange), and $\lambda=5$(red),
respectively.}
\label{VE-classical}
\end{figure}

Without loss of generality, let $\bar{q}=0$ and $\bar{p}=\sqrt{2M E_{0}}$. For convenience, let us introduce the dimensionless parameter $\lambda\equiv E_{0} /(k_B T)$ and set $V(t)=\frac{\langle (\Delta E)^{2}\rangle_{t}}{(k_B T)^{2}}$. We find that there exists a critical value of $\lambda$, denoted as $\lambda_{c}$, which is a function of $\alpha$, such that for $\lambda<\lambda_{c}$, $V(t)$ relaxes monotonically from 0 to 1, whereas for $\lambda>\lambda_{c}$, $V(t)$ evolves in an oscillatory way. In Fig.~\ref{VE-classical}, $V(t)$ for various initial energy values are shown for both the underdamped and the overdamped cases. In can be seen that, in both cases the $E$-variance could be highly nonmonotonic.

\begin{figure}[t]
\centering
\includegraphics[width=8cm]{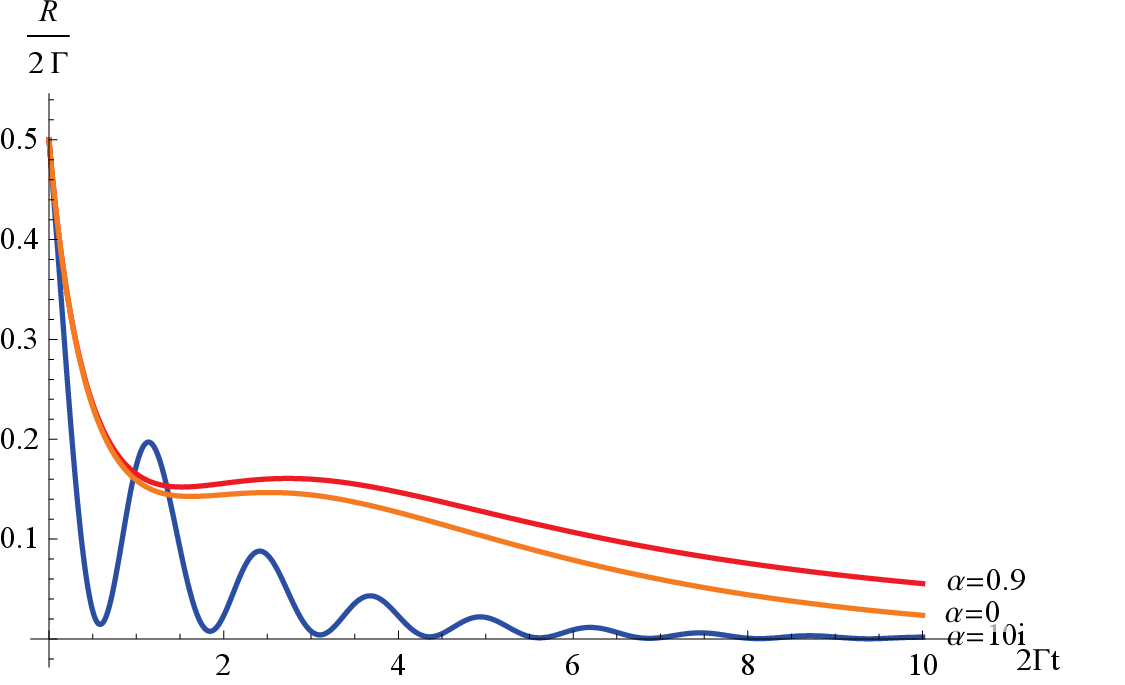}
\caption{Plot of the time derivative of the coarse-grained entropy
$R(t)\equiv {dS_{c}(t)}/{dt}$ for the classical harmonic oscillator when
it is underdamped with $\alpha=10i$ (blue), criticaldamped with $\alpha=0$
(orange), and overdamped with $\alpha=0.9$ (red), respectively.}
\label{Rt-classical}
\end{figure}

Finally, let us discuss the entropy of the classical damped harmonic oscillator. Considering Eqs.~\eqref{eq52} and \eqref{eq55}, the entropy can be expressed as
\begin{eqnarray}
\label{eq58}
S(t)=-\ln[h\int dq\int dp \rho^{2}(q,p)]~~~~~~~~~~~~~\nonumber\\
~~~~~=\ln\frac{4\pi^{2}}{h}-\ln\int dx\int dy~e^{2 \mathbf{Re}[v(x,y,t)]}\nonumber\\
=\ln\frac{4\pi}{h}+\frac{1}{2}\ln[4\tilde a(t) \tilde b(t)-\tilde c^{2}(t)],~~~~
\end{eqnarray}
where $h$ is a constant independent of $t$. Inserting Eq.~\eqref{eq56} into Eq.~\eqref{eq58} and considering a specific value of $h$ for convenience, i.e., $h={4\pi k_B T}/{\Gamma}$, we have
\begin{equation}
\label{eq59}
S(t)=\frac{1}{2}\ln[\frac{4(2e^{-\Gamma t}+\alpha^{2}(1-e^{-\Gamma t})^{2}
-2e^{-\Gamma t}\cosh\alpha\Gamma t)}{\alpha^{2}(1-\alpha^{2})}].
\end{equation}
Note that $S(t)$ given by \eqref{eq59} does not depend on the bath temperature, but it has a pathological point at $t=0$ that does not appear in the quantum case. To avoid this defect, we turn to the coarse-grained probability distribution $\rho_{c}(q,p)$ instead, which can be written in the form of \cite{Gu90}
$$\rho_{c}(q,p)=\frac{1}{2\pi\sigma_{q}\sigma_{p}}\int dQ\int dP e^{-(\frac{Q^{2}} {2\sigma_{q}^{2}}+\frac{P^{2}}{2\sigma_{p}^{2}})}\rho(q-Q,p-P).$$
Setting the characteristic function of $\rho_{c}(q,p,t)$ as $e^{v_{c}(x,y,t)}$, we have
\begin{equation}
\label{eq510}
v_{c}(x,y,t)=v(x,y,t)-\frac{1}{2}(\sigma_{p}^{2}x^{2}+\sigma_{q}^{2}y^{2}),
\end{equation}
the coarse-grained entropy of the system thus reads
\begin{equation}
\label{eq511}
S_{c}(t)=\frac{1}{2}\ln[4(\tilde a(t)-\frac{1}{2}\sigma_{p}^{2})(\tilde b(t)
-\frac{1}{2}\sigma_{q}^{2})-\tilde c^{2}(t)].
\end{equation}
Considering that the general properties of the coarse-grained entropy should not be affected by the coarse-graining details, for our aim here it is convenient to adopt a certain specific value of the coarse-graining size. In particular, by setting $\sigma_{q}^{2}=\frac{k_B T}{M\Gamma^{2}}$ and $\sigma_{p}^{2}=Mk_B T$, we obtain that
\begin{eqnarray}
\label{eq512}
S_{c}(t)=\frac{1}{2}\ln[\frac{2(5-\alpha^{2})}{1-\alpha^{2}}+
\frac{13-\alpha^{2}}{\alpha^{2}}e^{-\Gamma t}+
\frac{4}{1-\alpha^{2}}e^{-2\Gamma t}+\nonumber\\
e^{-\Gamma t}\frac{(\alpha^{2}-13)\cosh\alpha\Gamma t-\alpha(3+\alpha^{2})
\sinh\alpha\Gamma t}{\alpha^{2}(1-\alpha^{2})}],~
\end{eqnarray}
from which it follows that
\begin{equation}
\label{eq513}
S_{c}(0)=0;~~S_{c}(t)|_{t\rightarrow \infty}=\frac{1}{2}\ln
\frac{2(5-\alpha^{2})}{1-\alpha^{2}}.
\end{equation}
Moreover, by further analyzing $S(t)$ in detail, we find that the coarse-grained entropy $S_{c}(t)$ is actually a non-decreasing function and satisfies the requirement of the generalized $H$-theorem. As an illustration, in Fig.~\ref{Rt-classical} the time derivative of $S_c(t)$, $R(t)\equiv {dS_{c}(t)}/{dt}$, is displayed for the cases when the oscillator is underdamped, criticaldamped, and overdamped, respectively. Note that in all the three cases, $R(t)>0$ for a finite $t$ and $R(t)\to 0$ as $t\to \infty$, suggesting that $S_c(t)$ relaxes to its equilibrium value monotonically from below.

\section{Summary and Discussions}

We have demonstrated that, starting from a pure quantum state, the evolution of the harmonic oscillator governed by the quantum-optical master equation may have some unusual features in the course of thermalization.

The main findings by studying the entropy $S=-\ln{\textrm{Tr}_{\mathfrak{H}} [\rho^2]}$ are:
\emph{I.~} For an initial Gaussian state, there exists a critical value of $N_\beta$, i.e., $N_{c}=\sinh^{2}r$ (or equivalently a critical temperature of the thermal bath), such that if $N_{\beta}=(e^{\beta\hbar\omega}-1)^{-1}>N_{c}$, the entropy will relax monotonically from 0 to its equilibrium value $S_{eq} =\ln(1+2N_{\beta})$, whereas when $N_{\beta}<N_{c}$, the entropy will increase first to $S_{max}(r)$ (an infinitely increasing function of the squeeze factor $r$), then approaches asymptotically to its equilibrium value $S_{eq}$;
\emph{II.} For an initial Fock state $|n\rangle$ with $n>0$, there are two critical values of $N_{\beta}$ instead, one is $n$ and another is $N_{c}=n+d_{n}$, where $d_{n}$ is a positive number. As a result, an additional new phase of $S(t)$ appears for $n<N_{\beta}<N_{c}$, where $S(t)$ is dominated by a pair of extreme points in between $S(t)$ decreases. \emph{III.} In addition, the entropy of the quantum oscillator does not depend explicitly on its frequency $\omega$,
in contrast with the classical case where $\omega$ plays a role and distinctive features for underdamping and overdamping cases manifest. Moreover, the classical entropy is independent of the bath temperature.

We would like to remark that, assuming the entropy definition $S=-\ln{\textrm{Tr}_{\mathfrak{H}}[\rho^2]}$ proposed by Prigogine \cite{Prigogine73} rather than the von Neumann entropy $S=-\textrm{Tr}_{\mathfrak{H}}[\rho\ln{\rho}]$ facilitates greatly our analytical calculations. But note that, unlike the latter, the Prigogine entropy may fail to reflect some aspects of entropy such as concavity and subadditivity \cite{Wenrl78} and as a result, it cannot provide the information of the correlations between the subsystem and the reservoir \cite{Esposito10}. However, as a function of state, the Prigogine entropy serves well as a measure of the purity or dispersity of a mixed state. For these two reasons, the Prigogine entropy is ideal for our aim here -- to probe the relaxation process from a pure state towards a mixed state.

Finally, we discuss two possible ways for determining the valid range of our theoretical predictions based on the quantum-optical master equation.

I. By comparison with the results predicted by more sophisticated master equations.

Extensive studies have been conducted in recent decades on refining the Lindblad master equation from various aspects in order to better capture the evolution of an open quantum system. For example, constructing the generalized master equation for describing the dynamics beyond the Markov approximation \cite{Vega17}; recovering positivity of the Redfield equation via the partial secular approximation \cite{Farina19}; correcting the Redfield equation by using the exact mean-force Gibbs state \cite{Becker22}. Further studies based on these generalized master equations should be helpful. If the found relaxation phases and the transition between them remain, then we will have more faith that they are true properties of the open quantum harmonic oscillator.

II. By comparison with the related experimental studies.

As presumably the most important model system in physics, the quantum harmonic oscillator has been realized for experimental study in various laboratory settings, ranging from trapped electrons, atoms and/or quantum particles around the minimum of a potential well, vibrations in molecules and phones in solids, to electromagnetic field in optical cavity and so on (see Ref.~\cite{Cao23} and references therein). Moreover, there are various effective ways for preparing a given quantum state for the harmonic oscillator in laboratory (see, e.g.,
Ref.~\cite{Home15}). Therefore, an experimental check of our theoretical
predictions should be accessible.

However, as entropy is not an observable, for comparison with experimental study it is advisable to extract information about the dynamical evolution from experimentally accessible quantities. In view of the recent progress in experimental studies on quantum non-demolition photon counting as well as the dynamical evolution of photon number distribution of a relaxing optical field (e.g., Refs.~\cite{Brune90, Guerlin07, Brune08}), we calculate analytically the time dependence of the photon number distribution in Appendix, in an attempt to facilitate a direct comparison with the experimental measurements. Particularly, the case of an initial Gaussian state -- a common scenario in quantum optics and thermodynamics -- is discussed in detail. We hope these discussions would be interesting enough to stimulate experimental investigations.

On the other hand, in the applicable range of the quantum-optical master equation, we hope our results could be useful for designing new experimental approaches to generating and manipulating the quantum state of the harmonic oscillator, in view of the potentially important application situations such as quantum sensing, quantum simulations, quantum information processing/storage, etc. Indeed, dissipation is not always a hindrance for maintaining a quantum state. Rather, it can be harnessed to generate the desired quantum states (see, e.g., Ref. \cite{Liu23}). The decay of entropy, or the increase of purity or entanglement in certain time interval as revealed by our study, might play a role to this end.

\section*{Acknowledgments}

This work is supported by the National Key R\&D Program of China (Grant No. 2023YFA1407100) and the National Natural Science Foundation of China (Grant Nos. 12247106 and 12247101).

\appendix
\section*{Appendix: Evolution of diagonal elements of the density matrix in the Fock state basis}

To facilitate the comparison between the theoretical predictions from the master equation and the experimental results, we discuss the evolution of the diagonal elements of the density operator in the Fock state basis. The GCF proves equally valuable for this task, building on its prior utility in related analyses.

We denote the state of a quantum optical field by the density matrix $\rho(t)$ and set its GCF as $\varphi(g)=e^{iz \hbar+v(u,t)}$. According to Eq.~\eqref{eq23}, the diagonal elements of $\rho(t)$ in the Fock state basis can be written as
\begin{equation}
\label{eqA1}
P(n,t)\equiv \langle n|\rho(t)|n\rangle=\frac{1}{\pi}\int d^{2}u e^{v(u,t)
+\overline{v_{n}(u)}},
\end{equation}
where  $v_{n}(u)$ is given by Eq.~\eqref{eq42}. Since for the thermal state we have $v_{eq}(u)=-\frac{1}{2}(1+2N_{\beta})|u|^{2}$, it is easy to verify that
$$P(n,t)|_{t->\infty}=\frac{1}{\pi}\int d^{2}u e^{v_{eq}(u)+\overline{v_{n}(u)}}
=\frac{N_{\beta}^{n}}{(1+N_{\beta})^{n+1}}.$$

When the initial state is a Gaussian state (given by Eq.~\eqref{eq31}) and the dynamical evolution of the system is governed by Eq.~\eqref{eq11}, we may use the following transformations
\begin{equation}
\label{eqA2}
\bar{q}=2\sigma_{c}\alpha_{1},~~ \bar{p}=\frac{\hbar\alpha_{2}}{\sigma_{c}},~~
x=-\frac{2\sigma_{c}u_{1}}{\hbar},~~ y=\frac{u_{2}}{\sigma_{c}},~~\text{and}~\sigma
=\sigma_{c}e^{-r}
\end{equation}
to rewrite Eqs.~\eqref{eq32} and \eqref{eq33} as
\begin{equation}
\label{eqA3}
v(u,0)=2i(\alpha_{1}u_{2}-\alpha_{2}u_{1})-\frac{1}{2}(e^{2r}u_{1}^{2}
+e^{-2r}u_{2}^{2})
\end{equation}
and
\begin{equation}
\label{eqA4}
v(u,t)=A_{u}(t)u_{1}+B_{u}(t)u_{2}+a_{u}(t)u_{1}^{2}+b_{u}(t)u_{2}^{2}
+c_{u}(t)u_{1}u_{2},
\end{equation}
where
\begin{equation}
\label{eqA5}
\left\{\begin{array}{l}
A_{u}(t)=-2ie^{-\Gamma t}(\alpha_{2}\cos\omega t-\alpha_{1}\sin\omega t),\\
B_{u}(t)=2ie^{-\Gamma t}(\alpha_{1}\cos\omega t+\alpha_{2}\sin\omega t),\\
a_{u}(t)=-\frac{1}{2}e^{-2\Gamma t}((e^{2\Gamma t}-1)(2N_{\beta}+1)+e^{2r}
\cos^{2}\omega t+e^{-2r}\sin^{2}\omega t),\\
b_{u}(t)=-\frac{1}{2}e^{-2\Gamma t}((e^{2\Gamma t}-1)(2N_{\beta}+1)+e^{-2r}
\cos^{2}\omega t+e^{2r}\sin^{2}\omega t),\\
c_{u}(t)=e^{-2\Gamma t}\sinh2r\sin2\omega t.
\end{array}\right.
\end{equation}

\begin{figure}[t!]
\centering
\includegraphics[width=8.2cm]{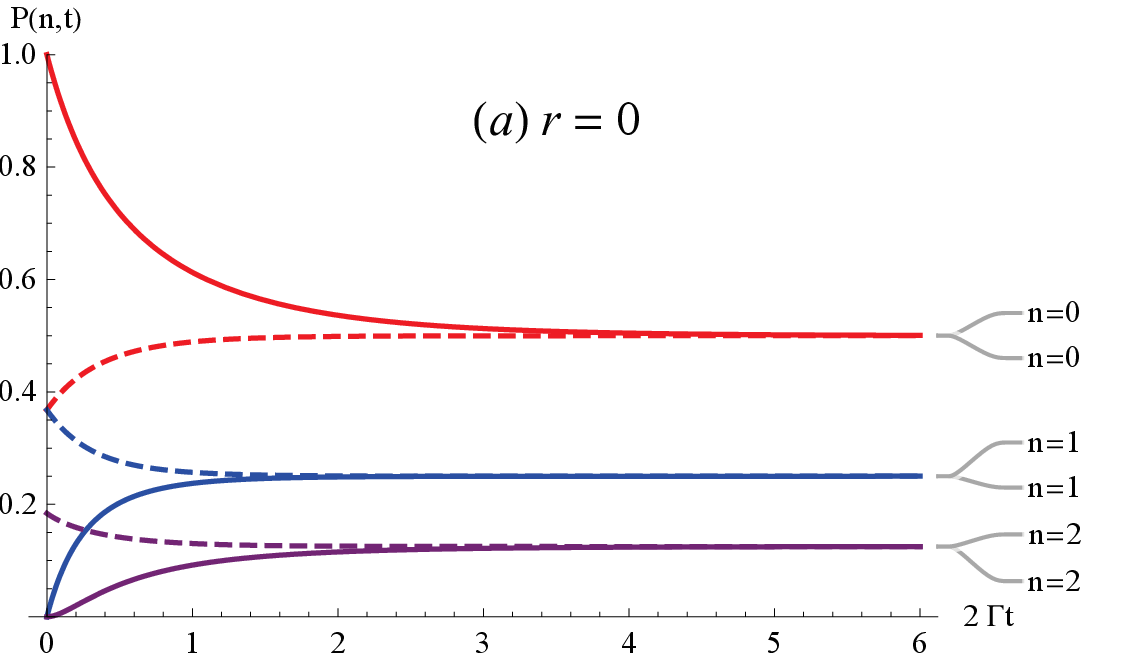}
\includegraphics[width=8.2cm]{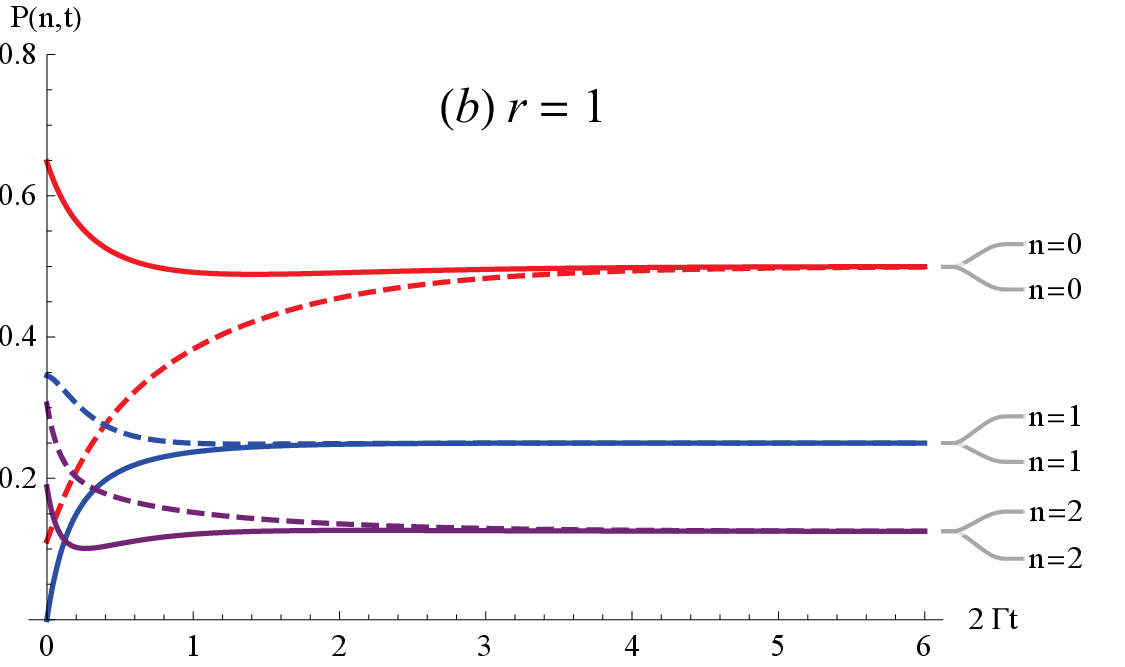}
\includegraphics[width=8.2cm]{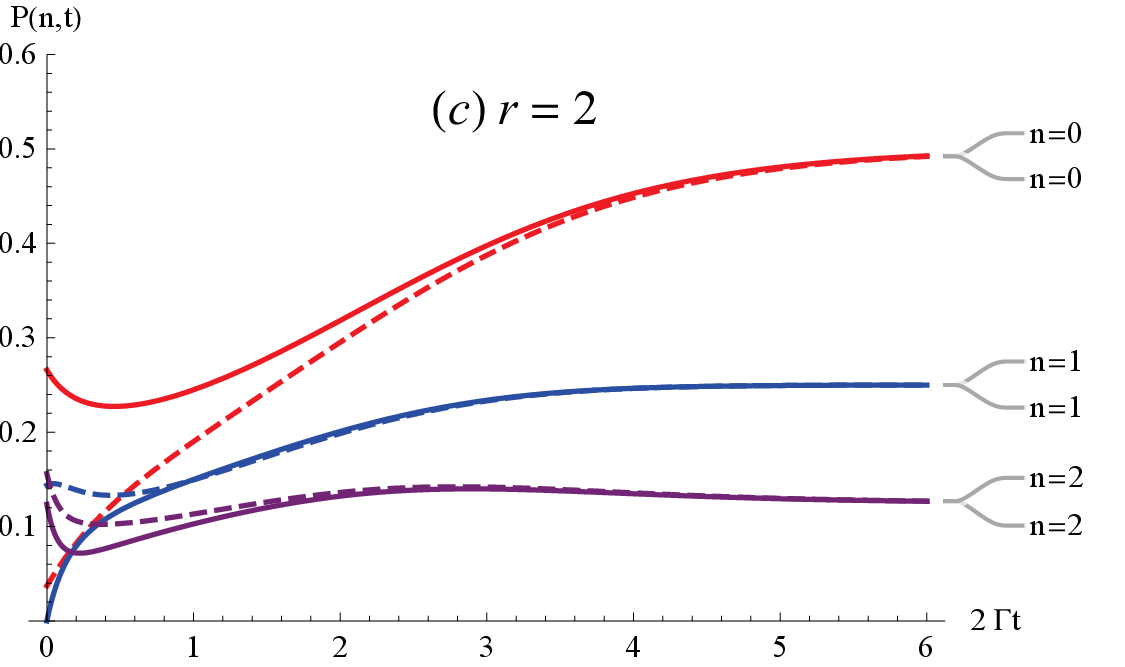}
\caption{Function $P(n,t)$ for an initial squeezed coherent state with
$N_{\beta}=1$ and $\alpha_{2}=0$. From top to bottom, the three panels
are for $r=0$ (a), $r=1$ (b), and $r=2$ (c), respectively. In all panels,
the solid (dashed) lines are for $\alpha_{1}=0$ ($\alpha_{1}=1$). The $n$
value for each curve is labeled at its right side.}
\label{Pnt}
\end{figure}

As $\langle n|[a^{\dag}a, \mathbf{\rho}]|n\rangle =0$, according to Eq.~\eqref{eq11}, $P(n,t)$ does not depend explicitly on the parameter $\omega$. To facilitate the calculation of $P(n,t)$, we can construct an ancillary density matrix ${\rho_{1}(t)}$ satisfying the master equation
\begin{equation}
\label{eqA6}
\frac{d{\rho_{1}}}{dt}=\Gamma(N_{\beta}+1)(2a\mathbf{\rho_{1}}a^{\dag}
-\{a^{\dag}a,{\rho_{1}}\})+\Gamma N_{\beta}(2a^{\dag}{\rho_{1}}a
-\{aa^{\dag},{\rho_{1}}\}),
\end{equation}
so that $P_{1}(n,t)\equiv \langle n|\rho_{1}(t)|n\rangle=P(n,t)$ if initially we have $P_{1}(n,0)=P(n,0)$.

Let the GCF of $\rho_{1}(t)$ be $\varphi_{1}(t,g)=e^{iz \hbar+v_{1}(u,t)}$. If initially $ v_{1}(u,0)= v(u,0)$, we have
\begin{equation}
\label{eqA7} v_{1}(u,t)=A_{1}(t)u_{1}+B_{1}(t)u_{2}+a_{1}(t)u_{1}^{2}
+b_{1}(t)u_{2}^{2}+c_{1}(t)u_{1}u_{2},
\end{equation}
where
\begin{equation}
\label{eqA8}
\left\{\begin{array}{l}
A_{1}(t)=-2ie^{-\Gamma t}\alpha_{2},\\
B_{1}(t)=2ie^{-\Gamma t}\alpha_{1},\\
a_{1}(t)=-\frac{1}{2}e^{-2\Gamma t}((e^{2\Gamma t}-1)(2N_{\beta}+1)+e^{2r}),\\
b_{1}(t)=-\frac{1}{2}e^{-2\Gamma t}((e^{2\Gamma t}-1)(2N_{\beta}+1)+e^{-2r}),\\
c_{1}(t)=0.
\end{array}\right.
\end{equation}
Thus we can rewrite Eq.~\eqref{eqA1} as
\begin{eqnarray}
\label{eqA9}
P(n,t)=\frac{1}{\pi}\int d^{2}u e^{v_{1}(u,t)+\overline{v_{n}(u)}}
=\frac{(-1)^{n}}{\pi 2^{2n}}\sum_{m=0}^{n}\frac{1}{m!(n-m)!}\times\nonumber\\
\int d u_{1}H_{2(n-m)}(u_{1})e^{A_{1}(t)u_{1}+(a_{1}(t)-\frac{1}{2})u_{1}^{2}}
\int d u_{2}H_{2m}(u_{2})e^{B_{1}(t)u_{2}+(b_{1}(t)-\frac{1}{2})u_{2}^{2}}.
\end{eqnarray}
Using the integral formula
\begin{equation}
\label{eqA10}
\int^{\infty}_{-\infty}dx H_{n}(x) e^{-a^{2}x^{2}+b x}=
\frac{\sqrt{\pi}(a^{2}-1)^{\frac{n}{2}}e^{\frac{b^{2}}{4a^{2}}}}
{a^{n+1}}H_{n}(\frac{b}{2a\sqrt{a^{2}-1}}),~~~\text{for}~a>0,
\end{equation}
we then have
\begin{eqnarray}
\label{eqA11}
P(n,t)=\frac{(-1)^{n}}{ 2^{2n}}e^{\frac{A_{1}(t)^{2}}{4d_{1}^{2}}
+\frac{B_{1}(t)^{2}}{4d_{2}^{2}}}\sum_{m=0}^{n}\frac{1}{m!(n-m)!}
\frac{(d_{1}^{2}-1)^{n-m}}{d_{1}^{2(n-m)+1}}
\frac{(d_{2}^{2}-1)^{m}}{d_{2}^{2m+1}}\times\nonumber\\
H_{2(n-m)}(\frac{A_{1}(t)}{2d_{1}\sqrt{d_{1}^{2}-1}})H_{2m}
(\frac{B_{1}(t)}{2d_{2}\sqrt{d_{2}^{2}-1}}),
\end{eqnarray}
where $d_{1}=\sqrt{\frac{1}{2}-a_{1}(t)}$ and $d_{2}=\sqrt{\frac{1}{2}-b_{1}(t)}$.

Equation~\eqref{eqA11} gives the evolution of the photon number distribution $P(n,t)$ based on the quantum-optical master equation \eqref{eq11}, where its dependence on the initial state parameter $\alpha$ and the squeeze factor $r$ is given explicitly. In Fig.~\ref{Pnt}, several cases for various parameter values are plotted, where the overall characteristics of $P(n,t)$ can be recognized. Note that an explicit expression of the generating function $G(z,t)$ of the photon number distribution $P(n,t)$ for an initial Gaussian state has been given in Ref.~\cite{Dodonov00} (see Eq.~(53) therein).

To show further the dependence of $P(n,t)$ on $N_{\beta}$, let us discuss a simple case: the relaxation process of $P(0,t)$ (i.e., $n=0$) for a squeezed vacuum. According to Eq.~\eqref{eqA11}, for a squeezed vacuum we have
\begin{equation}
\label{eqA12}
P(0,t)=\frac{1}{\sqrt{\frac{1}{2}-a_{1}(t)}\sqrt{\frac{1}{2}-b_{1}(t)}}.
\end{equation}

The time dependence of $P(0,t)$ can be captured by zeros of equation $R(t)={dP(0,t)}/{dt}$. It is found that the only finite solution is
\begin{equation}
\label{eqA13}
t_{m}=\frac{1}{2\Gamma}\ln[\frac{1+2N_{\beta}+2N_{\beta}^{2}-(1+2N_{\beta})
\cosh2r}{2(1+N_{\beta})(N_{\beta}-\sinh^{2} r)}].
\end{equation}
Therefore, when $N_{\beta}$ increases from 0 to $N_{c}= \sinh^{2} r$, $t_{m}$ increases from 0 to $\infty$. Note that $R(0)\leq 0 $, $P(0,t)$ bends downward if and only if $N_{\beta}< \sinh^{2} r$, as shown in Fig.~\ref{Pnt}. It is interesting to note that for this special case, the critical value of $N_{\beta}$ is the same as that given in Sec.~3 for the relaxation of entropy. However, for the cases where $n\neq 0$ and/or $\alpha\neq 0$, the dependence of $P(n,t)$ on $N_{\beta}$ becomes quite complicated.

\bibliography{QHO-arXiv-2nd}

\begin{thebibliography}{38}
\expandafter\ifx\csname natexlab\endcsname\relax\def\natexlab#1{#1}\fi
\expandafter\ifx\csname bibnamefont\endcsname\relax
  \def\bibnamefont#1{#1}\fi
\expandafter\ifx\csname bibfnamefont\endcsname\relax
  \def\bibfnamefont#1{#1}\fi
\expandafter\ifx\csname citenamefont\endcsname\relax
  \def\citenamefont#1{#1}\fi
\expandafter\ifx\csname url\endcsname\relax
  \def\url#1{\texttt{#1}}\fi
\expandafter\ifx\csname urlprefix\endcsname\relax\def\urlprefix{URL }\fi
\providecommand{\bibinfo}[2]{#2}
\providecommand{\eprint}[2][]{\url{#2}}

\bibitem[{\citenamefont{Breuer and Petruccione}(2002)}]{Breuer02}
\bibinfo{author}{\bibfnamefont{H.~P.} \bibnamefont{Breuer}} \bibnamefont{and}
  \bibinfo{author}{\bibfnamefont{F.}~\bibnamefont{Petruccione}},
  \emph{\bibinfo{title}{The Theory of Open Quantum Systems}}
  (\bibinfo{publisher}{Oxford University Press}, \bibinfo{year}{2002}).

\bibitem[{\citenamefont{Weiss}(2000)}]{Weiss00}
\bibinfo{author}{\bibfnamefont{D.}~\bibnamefont{Weiss}},
  \emph{\bibinfo{title}{Quantum Dissipative Systems}}
  (\bibinfo{publisher}{World Scientific}, \bibinfo{year}{2000}).

\bibitem[{\citenamefont{Louisell}(1973)}]{Louisell73}
\bibinfo{author}{\bibfnamefont{W.~H.} \bibnamefont{Louisell}},
  \emph{\bibinfo{title}{Quantum Statistical Properties of Radiation}}
  (\bibinfo{publisher}{John Wiley \& Sons}, \bibinfo{year}{1973}).

\bibitem[{\citenamefont{Haake}(1973)}]{Haake73}
\bibinfo{author}{\bibfnamefont{F.}~\bibnamefont{Haake}}, in
  \emph{\bibinfo{booktitle}{Springer Tracts in Modern Physics (Vol. 66)}}
  (\bibinfo{publisher}{Springer}, \bibinfo{year}{1973}).

\bibitem[{\citenamefont{Walls and Milburn}(2008)}]{Walls08}
\bibinfo{author}{\bibfnamefont{D.~F.} \bibnamefont{Walls}} \bibnamefont{and}
  \bibinfo{author}{\bibfnamefont{G.~J.} \bibnamefont{Milburn}},
  \emph{\bibinfo{title}{Quantum Optics}} (\bibinfo{publisher}{Springer},
  \bibinfo{year}{2008}).

\bibitem[{\citenamefont{Miguel}(2007)}]{Miguel07}
\bibinfo{author}{\bibfnamefont{O.}~\bibnamefont{Miguel}},
  \emph{\bibinfo{title}{Quantum Optics}} (\bibinfo{publisher}{Springer},
  \bibinfo{year}{2007}).

\bibitem[{\citenamefont{Carmichael}(2002)}]{Carmichael02}
\bibinfo{author}{\bibfnamefont{H.~J.} \bibnamefont{Carmichael}},
  \emph{\bibinfo{title}{Statistical Methods in Quantum Optics 1}}
  (\bibinfo{publisher}{Springer}, \bibinfo{year}{2002}).

\bibitem[{\citenamefont{Glauber}(1963)}]{Glauber63}
\bibinfo{author}{\bibfnamefont{R.~J.} \bibnamefont{Glauber}},
  \bibinfo{journal}{Phys. Rev.} \textbf{\bibinfo{volume}{131}},
  \bibinfo{pages}{2776} (\bibinfo{year}{1963}).

\bibitem[{\citenamefont{Sudarshan}(1963)}]{Sudarshan63}
\bibinfo{author}{\bibfnamefont{E.~C.~G.} \bibnamefont{Sudarshan}},
  \bibinfo{journal}{Phys. Rev. Lett.} \textbf{\bibinfo{volume}{10}},
  \bibinfo{pages}{277} (\bibinfo{year}{1963}).

\bibitem[{\citenamefont{Wigner}(1932)}]{Wigner32}
\bibinfo{author}{\bibfnamefont{E.}~\bibnamefont{Wigner}},
  \bibinfo{journal}{Phys. Rev.} \textbf{\bibinfo{volume}{40}},
  \bibinfo{pages}{749} (\bibinfo{year}{1932}).

\bibitem[{\citenamefont{Agarwal and Wolf}(1970)}]{Agarwal70}
\bibinfo{author}{\bibfnamefont{G.~S.} \bibnamefont{Agarwal}} \bibnamefont{and}
  \bibinfo{author}{\bibfnamefont{E.}~\bibnamefont{Wolf}},
  \bibinfo{journal}{Phys. Rev. D} \textbf{\bibinfo{volume}{2}},
  \bibinfo{pages}{2187} (\bibinfo{year}{1970}).

\bibitem[{\citenamefont{Risken}(1989)}]{Risken89}
\bibinfo{author}{\bibfnamefont{H.}~\bibnamefont{Risken}},
  \emph{\bibinfo{title}{The Fokker Planck Equation (see Appendix A4)}}
  (\bibinfo{publisher}{Springer}, \bibinfo{year}{1989}).

\bibitem[{\citenamefont{von Mises}(1964)}]{Mises64}
\bibinfo{author}{\bibfnamefont{R.}~\bibnamefont{von Mises}},
  \emph{\bibinfo{title}{Mathematical Theory of Probability and Statistics}}
  (\bibinfo{publisher}{Academic Press}, \bibinfo{year}{1964}).

\bibitem[{\citenamefont{Wang and Uhlenbeck}(1945)}]{Wang45}
\bibinfo{author}{\bibfnamefont{M.~C.} \bibnamefont{Wang}} \bibnamefont{and}
  \bibinfo{author}{\bibfnamefont{G.~E.} \bibnamefont{Uhlenbeck}},
  \bibinfo{journal}{Rev. Mod. Phys.} \textbf{\bibinfo{volume}{17}},
  \bibinfo{pages}{323} (\bibinfo{year}{1945}).

\bibitem[{\citenamefont{Dodonov et~al.}(2000)\citenamefont{Dodonov, Mizrahl,
  and de~Souza~Silva}}]{Dodonov00}
\bibinfo{author}{\bibfnamefont{V.~V.} \bibnamefont{Dodonov}},
  \bibinfo{author}{\bibfnamefont{S.~S.} \bibnamefont{Mizrahl}},
  \bibnamefont{and} \bibinfo{author}{\bibfnamefont{A.~L.}
  \bibnamefont{de~Souza~Silva}}, \bibinfo{journal}{J. Opt. B: Quantum
  Semiclass. Opt.} \textbf{\bibinfo{volume}{2}}, \bibinfo{pages}{271}
  (\bibinfo{year}{2000}).

\bibitem[{\citenamefont{Valeriano and Dodonov}(2020)}]{Valeriano20}
\bibinfo{author}{\bibfnamefont{J.~P.} \bibnamefont{Valeriano}}
  \bibnamefont{and} \bibinfo{author}{\bibfnamefont{V.~V.}
  \bibnamefont{Dodonov}}, \bibinfo{journal}{Phys. Lett. A}
  \textbf{\bibinfo{volume}{384}}, \bibinfo{pages}{126370}
  (\bibinfo{year}{2020}).

\bibitem[{\citenamefont{Marian and Marian}(1993{\natexlab{a}})}]{Marian93A}
\bibinfo{author}{\bibfnamefont{P.}~\bibnamefont{Marian}} \bibnamefont{and}
  \bibinfo{author}{\bibfnamefont{T.~A.} \bibnamefont{Marian}},
  \bibinfo{journal}{Phys. Rev. A} \textbf{\bibinfo{volume}{47}},
  \bibinfo{pages}{4474} (\bibinfo{year}{1993}{\natexlab{a}}).

\bibitem[{\citenamefont{Marian and Marian}(1993{\natexlab{b}})}]{Marian93B}
\bibinfo{author}{\bibfnamefont{P.}~\bibnamefont{Marian}} \bibnamefont{and}
  \bibinfo{author}{\bibfnamefont{T.~A.} \bibnamefont{Marian}},
  \bibinfo{journal}{Phys. Rev. A} \textbf{\bibinfo{volume}{47}},
  \bibinfo{pages}{4487} (\bibinfo{year}{1993}{\natexlab{b}}).

\bibitem[{\citenamefont{Marian et~al.}(2013)\citenamefont{Marian, Ghiu, and
  Marian}}]{Marian13}
\bibinfo{author}{\bibfnamefont{P.}~\bibnamefont{Marian}},
  \bibinfo{author}{\bibfnamefont{I.}~\bibnamefont{Ghiu}}, \bibnamefont{and}
  \bibinfo{author}{\bibfnamefont{T.~A.} \bibnamefont{Marian}},
  \bibinfo{journal}{Phys. Rev. A} \textbf{\bibinfo{volume}{88}},
  \bibinfo{pages}{012316} (\bibinfo{year}{2013}).

\bibitem[{\citenamefont{Gu}(1985)}]{Gu85}
\bibinfo{author}{\bibfnamefont{Y.}~\bibnamefont{Gu}}, \bibinfo{journal}{Phys.
  Rev. A} \textbf{\bibinfo{volume}{32}}, \bibinfo{pages}{1310}
  (\bibinfo{year}{1985}).

\bibitem[{\citenamefont{Gu}(1992)}]{Gu92}
\bibinfo{author}{\bibfnamefont{Y.}~\bibnamefont{Gu}}, \bibinfo{journal}{Science
  in China A} \textbf{\bibinfo{volume}{35}}, \bibinfo{pages}{200}
  (\bibinfo{year}{1992}).

\bibitem[{\citenamefont{Gu}(2020)}]{Gu20}
\bibinfo{author}{\bibfnamefont{Y.}~\bibnamefont{Gu}}, \bibinfo{journal}{Sci.
  China-Phys. Mech. Astron.} \textbf{\bibinfo{volume}{50}},
  \bibinfo{pages}{070002} (\bibinfo{year}{2020}).

\bibitem[{\citenamefont{Wolf}(1975)}]{Wolf75}
\bibinfo{author}{\bibfnamefont{K.~B.} \bibnamefont{Wolf}}, in
  \emph{\bibinfo{booktitle}{Group Theory and Its Applications (Vol. III, p.
  189)}} (\bibinfo{publisher}{Academic Press}, \bibinfo{year}{1975}).

\bibitem[{\citenamefont{Prigogine}(1973)}]{Prigogine73}
\bibinfo{author}{\bibfnamefont{I.}~\bibnamefont{Prigogine}}, in
  \emph{\bibinfo{booktitle}{The Physicist's Conception of Nature (p. 579)}}
  (\bibinfo{publisher}{Reidel}, \bibinfo{year}{1973}).

\bibitem[{\citenamefont{Zurek et~al.}(1993)\citenamefont{Zurek, Habib, and
  Paz}}]{Zurek93}
\bibinfo{author}{\bibfnamefont{W.~H.} \bibnamefont{Zurek}},
  \bibinfo{author}{\bibfnamefont{S.}~\bibnamefont{Habib}}, \bibnamefont{and}
  \bibinfo{author}{\bibfnamefont{J.~P.} \bibnamefont{Paz}},
  \bibinfo{journal}{Phys. Rev. Lett.} \textbf{\bibinfo{volume}{70}},
  \bibinfo{pages}{1187} (\bibinfo{year}{1993}).

\bibitem[{\citenamefont{Gradshteyn and Ryzhik}(1980)}]{Gradshteyn80}
\bibinfo{author}{\bibfnamefont{L.~S.} \bibnamefont{Gradshteyn}}
  \bibnamefont{and} \bibinfo{author}{\bibfnamefont{L.~M.}
  \bibnamefont{Ryzhik}}, \emph{\bibinfo{title}{Table of Integrals, Series and
  Products (see formula 7.374-5, p. 837)}} (\bibinfo{publisher}{Academic
  Press}, \bibinfo{year}{1980}).

\bibitem[{\citenamefont{Gu}(1990)}]{Gu90}
\bibinfo{author}{\bibfnamefont{Y.}~\bibnamefont{Gu}}, \bibinfo{journal}{Phys.
  Lett. A} \textbf{\bibinfo{volume}{149}}, \bibinfo{pages}{95}
  (\bibinfo{year}{1990}).

\bibitem[{\citenamefont{Wenrl}(1978)}]{Wenrl78}
\bibinfo{author}{\bibfnamefont{A.}~\bibnamefont{Wenrl}}, \bibinfo{journal}{Rev.
  Mod. Phys.} \textbf{\bibinfo{volume}{50}}, \bibinfo{pages}{221}
  (\bibinfo{year}{1978}).

\bibitem[{\citenamefont{Esposito et~al.}(2010)\citenamefont{Esposito,
  Lindenberg, and van~den Breck}}]{Esposito10}
\bibinfo{author}{\bibfnamefont{M.}~\bibnamefont{Esposito}},
  \bibinfo{author}{\bibfnamefont{K.}~\bibnamefont{Lindenberg}},
  \bibnamefont{and} \bibinfo{author}{\bibfnamefont{C.}~\bibnamefont{van~den
  Breck}}, \bibinfo{journal}{New J. Phys.} \textbf{\bibinfo{volume}{12}},
  \bibinfo{pages}{013013} (\bibinfo{year}{2010}).

\bibitem[{\citenamefont{de~Vega and Alonso}(2017)}]{Vega17}
\bibinfo{author}{\bibfnamefont{I.}~\bibnamefont{de~Vega}} \bibnamefont{and}
  \bibinfo{author}{\bibfnamefont{D.}~\bibnamefont{Alonso}},
  \bibinfo{journal}{Rev. Mod. Phys.} \textbf{\bibinfo{volume}{89}},
  \bibinfo{pages}{015001} (\bibinfo{year}{2017}).

\bibitem[{\citenamefont{Farina and Giovannetti}(2019)}]{Farina19}
\bibinfo{author}{\bibfnamefont{D.}~\bibnamefont{Farina}} \bibnamefont{and}
  \bibinfo{author}{\bibfnamefont{V.}~\bibnamefont{Giovannetti}},
  \bibinfo{journal}{Phys. Rev. A} \textbf{\bibinfo{volume}{100}},
  \bibinfo{pages}{012107} (\bibinfo{year}{2019}).

\bibitem[{\citenamefont{Becker et~al.}(2022)\citenamefont{Becker, Schnell, and
  Thingna}}]{Becker22}
\bibinfo{author}{\bibfnamefont{T.}~\bibnamefont{Becker}},
  \bibinfo{author}{\bibfnamefont{A.}~\bibnamefont{Schnell}}, \bibnamefont{and}
  \bibinfo{author}{\bibfnamefont{J.}~\bibnamefont{Thingna}},
  \bibinfo{journal}{Phys. Rev. Lett.} \textbf{\bibinfo{volume}{129}},
  \bibinfo{pages}{200403} (\bibinfo{year}{2022}).

\bibitem[{\citenamefont{Mei et~al.}(2023)\citenamefont{Mei, Suo, Mao, Feng, and
  Cao}}]{Cao23}
\bibinfo{author}{\bibfnamefont{G.}~\bibnamefont{Mei}},
  \bibinfo{author}{\bibfnamefont{P.}~\bibnamefont{Suo}},
  \bibinfo{author}{\bibfnamefont{L.}~\bibnamefont{Mao}},
  \bibinfo{author}{\bibfnamefont{M.}~\bibnamefont{Feng}}, \bibnamefont{and}
  \bibinfo{author}{\bibfnamefont{L.}~\bibnamefont{Cao}},
  \bibinfo{journal}{Frontiers of Physics} \textbf{\bibinfo{volume}{18}},
  \bibinfo{pages}{13310} (\bibinfo{year}{2023}).

\bibitem[{\citenamefont{Kienzler et~al.}(2015)\citenamefont{Kienzler, Lo,
  Keitch, de~Clercq, Leupold, Lindenfelser, Marinelli, Negnevitsky, and
  Home}}]{Home15}
\bibinfo{author}{\bibfnamefont{D.}~\bibnamefont{Kienzler}},
  \bibinfo{author}{\bibfnamefont{H.-Y.} \bibnamefont{Lo}},
  \bibinfo{author}{\bibfnamefont{B.}~\bibnamefont{Keitch}},
  \bibinfo{author}{\bibfnamefont{L.}~\bibnamefont{de~Clercq}},
  \bibinfo{author}{\bibfnamefont{F.}~\bibnamefont{Leupold}},
  \bibinfo{author}{\bibfnamefont{F.}~\bibnamefont{Lindenfelser}},
  \bibinfo{author}{\bibfnamefont{M.}~\bibnamefont{Marinelli}},
  \bibinfo{author}{\bibfnamefont{V.}~\bibnamefont{Negnevitsky}},
  \bibnamefont{and} \bibinfo{author}{\bibfnamefont{J.~P.} \bibnamefont{Home}},
  \bibinfo{journal}{Science} \textbf{\bibinfo{volume}{347}},
  \bibinfo{pages}{53} (\bibinfo{year}{2015}).

\bibitem[{\citenamefont{Brune et~al.}(1990)\citenamefont{Brune, Haroche,
  Lefevre, Raimond, and Zagury}}]{Brune90}
\bibinfo{author}{\bibfnamefont{M.}~\bibnamefont{Brune}},
  \bibinfo{author}{\bibfnamefont{S.}~\bibnamefont{Haroche}},
  \bibinfo{author}{\bibfnamefont{V.}~\bibnamefont{Lefevre}},
  \bibinfo{author}{\bibfnamefont{J.~M.} \bibnamefont{Raimond}},
  \bibnamefont{and} \bibinfo{author}{\bibfnamefont{N.}~\bibnamefont{Zagury}},
  \bibinfo{journal}{Phys. Rev. Lett.} \textbf{\bibinfo{volume}{65}},
  \bibinfo{pages}{976} (\bibinfo{year}{1990}).

\bibitem[{\citenamefont{Guerlin et~al.}(2007)\citenamefont{Guerlin, Bernu,
  Deleglise, Sayrin, Gleyzes, Kuhr, Brune, Raimond, and Haroche}}]{Guerlin07}
\bibinfo{author}{\bibfnamefont{C.}~\bibnamefont{Guerlin}},
  \bibinfo{author}{\bibfnamefont{J.}~\bibnamefont{Bernu}},
  \bibinfo{author}{\bibfnamefont{S.}~\bibnamefont{Deleglise}},
  \bibinfo{author}{\bibfnamefont{C.}~\bibnamefont{Sayrin}},
  \bibinfo{author}{\bibfnamefont{S.}~\bibnamefont{Gleyzes}},
  \bibinfo{author}{\bibfnamefont{S.}~\bibnamefont{Kuhr}},
  \bibinfo{author}{\bibfnamefont{M.}~\bibnamefont{Brune}},
  \bibinfo{author}{\bibfnamefont{J.~M.} \bibnamefont{Raimond}},
  \bibnamefont{and} \bibinfo{author}{\bibfnamefont{S.}~\bibnamefont{Haroche}},
  \bibinfo{journal}{Nature} \textbf{\bibinfo{volume}{448}},
  \bibinfo{pages}{889} (\bibinfo{year}{2007}).

\bibitem[{\citenamefont{Brune et~al.}(2008)\citenamefont{Brune, Bernu, Guerlin,
  Deleglise, Sayrin, Gleyzes, Kuhr, Dotsenko, Raimond, and Haroche}}]{Brune08}
\bibinfo{author}{\bibfnamefont{M.}~\bibnamefont{Brune}},
  \bibinfo{author}{\bibfnamefont{J.}~\bibnamefont{Bernu}},
  \bibinfo{author}{\bibfnamefont{C.}~\bibnamefont{Guerlin}},
  \bibinfo{author}{\bibfnamefont{S.}~\bibnamefont{Deleglise}},
  \bibinfo{author}{\bibfnamefont{C.}~\bibnamefont{Sayrin}},
  \bibinfo{author}{\bibfnamefont{S.}~\bibnamefont{Gleyzes}},
  \bibinfo{author}{\bibfnamefont{S.}~\bibnamefont{Kuhr}},
  \bibinfo{author}{\bibfnamefont{I.}~\bibnamefont{Dotsenko}},
  \bibinfo{author}{\bibfnamefont{J.~M.} \bibnamefont{Raimond}},
  \bibnamefont{and} \bibinfo{author}{\bibfnamefont{S.}~\bibnamefont{Haroche}},
  \bibinfo{journal}{Phys. Rev. Lett.} \textbf{\bibinfo{volume}{101}},
  \bibinfo{pages}{240402} (\bibinfo{year}{2008}).

\bibitem[{\citenamefont{Liu and Nie}(2023)}]{Liu23}
\bibinfo{author}{\bibfnamefont{J.}~\bibnamefont{Liu}} \bibnamefont{and}
  \bibinfo{author}{\bibfnamefont{H.}~\bibnamefont{Nie}},
  \bibinfo{journal}{Phys. Rev. A} \textbf{\bibinfo{volume}{107}},
  \bibinfo{pages}{052608} (\bibinfo{year}{2023}).

\end{thebibliography}

\end{document}